# Density Functional Theory of Material Design: Fundamentals and Applications - I

Prashant Singh[1] and Manoj K. Harbola[2,*]

[1]Ames Laboratory, U.S. Department of Energy, Iowa State University, Ames, Iowa 50011 USA

[2]Department of Physics, Indian Institute of Technology, Kanpur
Kanpur – 208016, India

**Abstract**

This article is part-I of a review of density-functional theory (DFT) that is the most widely used method for calculating electronic structure of materials. The accuracy and ease of numerical implementation of DFT methods has resulted in its extensive use for materials design and discovery and has thus ushered in the new field of computational material science. In this article we start with an introduction to Schrödinger equation and methods of its solutions. After presenting exact results for some well-known systems, difficulties encountered in solving the equation for interacting electrons are described. How these difficulties are handled using the variational principle for the energy to obtain approximate solutions of the Schrödinger equation is discussed. The resulting Hartree and Hartree-Fock theories are presented along with results they give for atomic and solid-state systems. We then describe Thomas-Fermi theory and its extensions which were the initial attempts to formulate many-electron problem in terms of electronic density of a system. Having described these theories, we introduce modern density functional theory by discussing Hohenberg-Kohn theorems that form its foundations. We then go on to discuss Kohn-Sham formulation of density-functional theory in its exact form. Next, local density approximation is introduced and solutions of Kohn-Sham equation for some representative systems, obtained using the local density approximation, are presented. We end part-I of the review describing the contents of part-II.

**Keywords:** *Schrödinger equation, Density functional theory, Kohn-Sham equation, Variational method, Hartree-Fock theory, Local-density approximation,*

**Corresponding author address**:
FB-476, Department of Physics Indian,
Institute of Technology Kanpur, Kanpur,
Uttar Pradesh, INDIA-208016
**Email**: mkh@iitk.ac.in

## 1. Introduction: Motivation for Density Functional Theory (DFT)

Properties of materials are determined by how their constituents – electrons and ions – respond to different stimuli applied to them. For example, bulk modulus of a system is determined by how much the underlying arrangement of the constituent atoms or molecules in the corresponding crystal change when it is subjected to pressure; colour of a substance depends on how its electrons change their state when light interacts with it. Similarly, electric properties of matter are based on the response of its electronic charge to an electric field and its magnetic properties on the sum of magnetic moments (orbital or spin) of electrons in it when subjected to a magnetic field. Thus, theoretical understanding of properties of different materials requires a knowledge of how the atoms, ions and electrons forming it are arranged and how their distribution responds to an external stimulant. The problem is thus solved in two stages: we first calculate the structure of a given system without any external influence and then build up on it to calculate the response of the system to an applied field. Both are described by the wavefunction of the system. The wavefunction in turn is obtained by solving the Schrödinger equation [1, 2]. Thus, an understanding of what a wavefunction represents and how properties of interest are calculated from it is important for a theoretical material scientist. Furthermore, the Schrödinger equation is a complicated differential equation and can be solved exactly only for a limited number of systems. It is therefore equally important to understand how its solution can be facilitated by making reasonable approximations or by some other methods, and how these are solved on computers numerically. These methods and the associated numerical techniques have been developed ever since the Schrödinger equation was discovered and have evolved to become more and more accurate with time. Among all these, the method of choice for material scientists is the subject matter of this article and is known as density-functional theory (DFT).

In the following we start with a discussion of the Schrödinger equation and the wavefunctions. We then present the difficulties faced in solving the many-particle Schrödinger equation and describe some methods of solving it to obtain approximate wavefunctions. This is followed by a brief description of the initial attempts made to bypass solving the Schrödinger equation for the wavefunction by recasting the problem in terms of the density. Next, we discuss how these seminal efforts culminate in the exact reformulation of many-body Schrödinger equation in terms of density and make it possible to apply it to a range of systems with equal ease. This to a large extent has also been aided by the increase in computational power, since the corresponding equations can be solved only numerically. Thus, numerical techniques form an essential component of any theory of material design and are discussed along with the formulation of DFT. The paper is concluded by presenting different possible directions that can be taken in applying DFT to obtain properties of material.

### 1a. Many-electron Schrödinger equation

Materials of interest in this article are made up of positively charged nuclei or ions of atoms and electrons. Consider a neutral system that has $N_I$ ions and $N$ electrons with their masses being $M$ (for simplicity we have assumed that all the ions have the same mass) and $m$, respectively. The position of ions is represented by $\boldsymbol{R}_i (i = 1 \cdots N_I)$ and those of electrons by $\boldsymbol{r}_i (i = 1 \cdots N)$. The corresponding stationary-state wavefunction $\Psi(\boldsymbol{R}_1 \cdots \boldsymbol{R}_{N_I}; \boldsymbol{r}_1 \cdots \boldsymbol{r}_N)$ is a function of both the nuclear and the electron coordinates. It is a solution of the time-independent Schrödinger equation

$$H\Psi = E\Psi, \tag{1}$$

where

$$H = -\sum_{i=1}^{N_I} \frac{\hbar^2}{2M} \nabla_{\boldsymbol{R}_i}^2 - \sum_{i=1}^{N} \frac{\hbar^2}{2m} \nabla_{\boldsymbol{r}_i}^2 + V_{II}(\boldsymbol{R}_1 \cdots \boldsymbol{R}_{N_I}) + \sum_{i=1}^{N} v_{ext}(\boldsymbol{R}_1 \cdots \boldsymbol{R}_{N_I}; \boldsymbol{r}_i)$$

$$+ \frac{1}{2} \frac{e^2}{4\pi\epsilon_0} \sum_{\substack{i,j \\ i \neq j}} \frac{1}{|\boldsymbol{r}_i - \boldsymbol{r}_j|} \tag{2}$$

with $\hbar$ being Planck's constant divided by $2\pi$, is the Hamiltonian operator and referred to simply as the Hamiltonian and contains second-order derivatives with respect to both sets of coordinates, i.e. $\{\boldsymbol{R}_i\}$ and $\{\boldsymbol{r}_i\}$. These derivatives represent the kinetic energy operators for the corresponding particles. Additionally, the Hamiltonian contains the potential energy of the system (as operators, these are multiplicative operators in contrast to the kinetic energy operator that is a differential operator) that consists of the energy of interaction $V_{II}(R_1 \cdots R_{N_I})$ between the ions, $v_{ext}(R_1 \cdots R_{N_I}; r_i)$ between the nuclei and the electrons (it will be referred to it as the external potential in the manuscript, as is done in the literature) and the last term which is the energy of interaction between the electrons. Notice that because of the external potential, the motion of the constituent ions and the electrons is coupled. The first simplification in the equation comes by decoupling this motion. The simplest way to do it is to make the **static approximation** [3] and take the ionic positions to be fixed at their equilibrium positions $\{R_i^0\}$ in the system – this effectively amounts to taking the mass of the ions to be infinitely large. Then their kinetic energy vanishes and the corresponding potential energy $V_{II}(R_1 \cdots R_{N_I})$ is a constant. As a result, only the electronic degree of freedom is to be considered and one solves the Schrödinger equation for the Hamiltonian

$$H_{el} = -\sum_{i=1}^{N} \frac{\hbar^2}{2m} \nabla_{r_i}^2 + \sum_{i=1}^{N} v_{ext}(R_1^0 \cdots R_{N_I}^0; r_i) + \frac{1}{2} \frac{e^2}{4\pi\epsilon_0} \sum_{\substack{i,j \\ i \neq j}} \frac{1}{|r_i - r_j|} \qquad (3)$$

Using the electronic wavefunctions $\psi(R_1^0 \cdots R_{N_I}^0; r_1 \cdots r_N)$ so obtained, one can then build up the complete wavefunction as a product

$$\Psi(R_1 \cdots R_{N_I}; r_1 \cdots r_N) = \psi(R_1^0 \cdots R_{N_I}^0; r_1 \cdots r_N)\chi(R_1 \cdots R_{N_I}) \qquad (4)$$

of the electronic and the ionic wavefunctions $\chi(R_1 \cdots R_{N_I})$. The wave-equation for the latter is obtained by substituting the wavefunction of Eq. (4) in the full Schrödinger equation of Eq. (1) and taking its expectation value with respect to $\psi(R_1^0 \cdots R_{N_I}^0; r_1 \cdots r_N)$. A variant of static approximation is the **Born-Oppenheimer approximation** or the **adiabatic approximation** [3] where the electronic wavefunction is calculated with the external potential determined by the instantaneous position of the ions and the total wavefunction is written as the product of the electronic function so determined and the ionic wavefunction. Again, the latter is obtained by solving the equation that is derived from the full Schrödinger equation of Eq. (1). The ionic and electronic wavefunctions can be separated because of the enormous difference between the masses of the ions and electrons (keep in mind that proton's mass is about 1800 times that of an electron). This is explained in the box below on the basis of the uncertainty principle.

---

*The adiabatic approximation*

Imagine a collection of electrons and ions which move over the length scale $l$. Then by the uncertainty principle, momentum of each of these is of the order of $\hbar/l$. However, since ions are of much larger mass than the electrons, variance of their speeds will be significantly smaller in comparison to those of electrons. Hence, as the zeroth level approximation, movement of ions about their centre of mass can be ignored and electronic movement can be decoupled from it.

---

In either case, one must first solve the electronic Schrödinger equation and then use the corresponding wavefunction to determine ionic motion; the latter is described approximately even if the electronic part is known exactly. For a critical discussion of the adiabatic approximation and its applicability, we refer the reader to ref. [4]. It suffices here to say that in applying the adiabatic approximation, motion of nuclei is restricted to small regions around their equilibrium position [4]. In this article, however, we will be concerned with solving the electronic problem assuming a given position of ions.

For a given position of ions, the Schrödinger equation for the electrons is

$$\left[-\sum_{i=1}^{N}\frac{1}{2}\nabla_{r_i}^2 + \sum_{i=1}^{N} v_{ext}(r_i) + \frac{1}{2}\sum_{\substack{i,j \\ i\neq j}}\frac{1}{|r_i - r_j|}\right]\psi(r_1\cdots r_N) = E\psi(r_1\cdots r_N) \ , \quad (5)$$

where for brevity, we have not shown the dependence of $v_{ext}(r_i)$ or the electronic wavefunction $\psi(r_1\cdots r_N)$ on the ion coordinates explicitly. Furthermore, we have also taken $\hbar = m = |e| = 1$ and $\frac{e^2}{4\pi\epsilon_0} = 1$. These are known as atomic units (a.u.), and the unit of length is 0.529A° in terms of these, which is the Bohr radius of an electron in hydrogen atom, and the unit of energy is 27.21 eV. Notice that if the electrons were not interacting, the last term in the square brackets in Eq. (5) will not be there. The wavefunction is required to satisfy certain mathematical properties [1, 2] related to its continuity and integrability. The equation is solved with appropriate boundary conditions for the wavefunction. These conditions and the required mathematical properties can be satisfied only for certain values of $E$. These are known as the eigenvalues for the energy and the corresponding wavefunctions as the eigenfunctions. We refer the reader to [1, 2] to review how the single-electron Schrödinger equation

$$\left[-\frac{1}{2}\nabla^2 + v_{ext}(r)\right]\psi(r) = E\psi(r) \qquad (6)$$

is solved analytically for some specific systems and numerically for general external potentials.

---

**A note before further discussion**

For brevity, right now we will take the wavefunction to depend only on the space coordinates and postpone the inclusion of spin in it to little later. The electronic wavefunction is antisymmetric with respect to the exchange of all coordinates (space and spin) of any two electrons. Therefore, if the space and spin components of a wavefunction can be separated, one of them will be symmetric and the other antisymmetric with respect to such an exchange.

---

The wavefunction $\psi(r_1\cdots r_N)$ is in general complex and its absolute square is proportional to the probability density $p(r_1\cdots r_N)$ of finding electrons at positions $(r_1\cdots r_N)$. Therefore, the wavefunctions is normalized so that

$$\int |\psi(r_1\cdots r_N)|^2 dr_1\cdots dr_N = 1 \ . \qquad (7)$$

and the probability density

$$p(r_1\cdots r_N) = |\psi(r_1\cdots r_N)|^2 \ . \qquad (8)$$

As a result, the density $\rho(r)$ of electrons is given as

$$\rho(r) = N\int |\psi(r_1 = r;\ r_2\cdots r_N)|^2 dr_2\cdots dr_N \ . \qquad (9)$$

In writing the expression for the density above, use has been made of the property of the space component of the wavefunction for electrons that it is either symmetric or antisymmetric, which makes $|\psi(r_1\cdots r_N)|^2$ symmetric, with respect to the exchange of coordinates of any two electrons. Any property of interest for a given system is represented by an operator $\hat{O}$ and is calculated from the wavefunction as the expectation value

$$\langle\psi|\hat{O}|\psi\rangle = \int \psi^*(\boldsymbol{r}_1 \cdots \boldsymbol{r}_N) \, \hat{O} \, \psi(\boldsymbol{r}_1 \cdots \boldsymbol{r}_N) d\boldsymbol{r}_1 \cdots d\boldsymbol{r}_N \tag{10}$$

of this operator; it is also denoted as $\langle\hat{O}\rangle$ in short. For example, the operator for the density is [5]

$$\hat{\rho}(\boldsymbol{r}) = \sum_{i=1}^{N} \delta(\boldsymbol{r} - \boldsymbol{r}_i) \quad . \tag{11}$$

It is easily verified that expectation value of $\hat{\rho}(\boldsymbol{r})$ leads to Eq. (9) for the density. Notice that the expectation value of an operator will come out to be equal to the eigenvalue if the wavefunction is an eigenfunction of the operator.

To illustrate various ideas discussed in this article, we will make use of two simplest many-electron systems which contain two electrons. In both of these, we consider the ground-state of electrons for which the space-dependent wavefunction is symmetric with respect to the position of the electrons. One of the examples is that of two electrons moving in a potential proportional to the square their distance from the origin. This is known as the Hookium atom. The Schrödinger equation for the Hookium [6, 7]

$$\left[-\frac{1}{2}\nabla_1^2 - \frac{1}{2}\nabla_2^2 + \frac{1}{2}\omega^2 r_1^2 + \frac{1}{2}\omega^2 r_2^2 + \frac{1}{|\boldsymbol{r}_1 - \boldsymbol{r}_2|}\right]\psi(\boldsymbol{r}_1, \boldsymbol{r}_2) = E\psi(\boldsymbol{r}_1, \boldsymbol{r}_2) \; . \tag{12}$$

is exactly solvable analytically [6] for $\omega = \frac{1}{2}$. On solving it, the energy of the system comes out to be exactly 2. The corresponding wavefunction is

$$\psi(\boldsymbol{r}_1, \boldsymbol{r}_2) = C_N \left(1 + \frac{1}{2}|\boldsymbol{r}_1 - \boldsymbol{r}_2|\right) e^{-\frac{1}{4}(r_1^2 + r_2^2)} \quad , \tag{13a}$$

where

$$C_N = \frac{\pi^{3/4}}{\sqrt{8 + 5\pi^{1/2}}} \quad . \tag{13b}$$

The expression for the density of the system can be worked out using Eq. (9) and is given in ref. [4]. Let us compare the wavefunction given in Eq. (13a) with that obtained if interaction between electrons was not there. In that case the wavefunction will be proportional to $e^{-\frac{1}{4}(r_1^2 + r_2^2)}$. Thus, the effect of interaction on the wavefunction is to make it larger as the distance $|\boldsymbol{r}_1 - \boldsymbol{r}_2|$ between the electrons increases, thereby increasing the relative probability of them being farther apart from each other. Evidently, for $r_1 = r_2$ electrons have much higher probability of being on the opposite sides of the origin than for being on the same side; if they were not interacting, these probabilities would be equal. The wavefunction thus keeps the electrons **correlated**.

The other example is that of the Helium atom, which has two electrons with a nucleus made of two protons and two neutrons. The corresponding Schrödinger equation satisfied by the electrons is

$$\left[-\frac{1}{2}\nabla_1^2 - \frac{1}{2}\nabla_2^2 - \frac{2}{r_1} - \frac{2}{r_2} + \frac{1}{|\boldsymbol{r}_1 - \boldsymbol{r}_2|}\right]\psi(\boldsymbol{r}_1, \boldsymbol{r}_2) = E\psi(\boldsymbol{r}_1, \boldsymbol{r}_2) \; . \tag{14}$$

This equation cannot be solved analytically. However, highly accurate numerical solutions [8-11] exist for it and we will take them to be equivalent to exact results.  To facilitate faster calculations, very accurate semi-analytic wavefunctions for these systems have also been developed over the years [12-14].

From the examples above, we see that even for the simplest systems, exact solution exists only for a specific case.  In general, exact solution of the many-electron Schrödinger equation (Eq. 5) cannot be obtained.  This difficulty arises because the electron-electron interaction energy term in the Schrödinger equation does not allow separation of variables.  As a result, the wavefunction cannot be written as a product of single particle wavefunctions, which are solutions of single particle Schrödinger equation like that of Eq. 6.  Furthermore, purely numerical solution may also not be possible because of the enormous requirement of memory.  Therefore, methods to obtain accurate solutions have to be developed to make further progress.  This is done using the variational principle for the energy and is described next.

**1b. The variational method of obtaining approximate solutions [2]**

The ground-state (the lowest energy state) of any system has the following property.  If the expectation value of the Hamiltonian for a system is taken with respect to a normalized wavefunction satisfying the appropriate boundary conditions, it will always be greater than or equal to the exact ground-state energy $E_0$ . Thus

$$\langle\psi|H|\psi\rangle \geq E_0 , \quad (15)$$

where the equality holds if the wavefunction is exact.  This property of the ground-state can be used to find an approximate ground-state wavefunction for the system.  The procedure for this is as follows.

First, we choose an approximate wavefunction appropriate for the given system.  The wavefunction can be chosen as a function with its analytical form appropriate for the system under consideration.  In the examples discussed below, we follow this path.   In a different approach, the wavefunction is written as a linear combination of a set of functions with the expansion coefficients treated as parameters of the wavefunction; the basis-functions employed for this are according to the system one is dealing with.  For example, in performing calculations on solids, often the wavefunction is expanded in terms of plane waves or related functions [15].  Having chosen a wavefunction, it is normalized and the expectation value $\langle H \rangle$ of the Hamiltonian is calculated using it. The wavefunction is then varied until $\langle H \rangle$ achieves its minimum value. This can be done in two following ways.  First, we can choose a functional form with some parameters in it that are varied to minimize $\langle H \rangle$.  Wavefunctions written as a linear combination automatically fall in this category.  Secondly, we can derive a variational equation for the wavefunction and solve it.  In either case, the wavefunction obtained is an approximation to the true ground-state wavefunction and the minimum expectation value gives approximate energy of the ground-state.  The energy obtained is always above the true ground-state energy and thus represents an upper bound to it.  A nice feature of the variational method is that it offers a way of systematically improving the wavefunction by including more and more parameters in it.  As this is done, the expectation value becomes lower and lower, and it approaches the exact value of the energy from above. Let us illustrate this by applying the method to the examples of two-electron systems considered earlier.

For two-electron systems, the simplest wavefunction that we can choose [2] is the ***product wavefunction*** of two single-electron wavefunctions $\varphi(\boldsymbol{r})$, i.e.:

$$\psi(\boldsymbol{r}_1,\boldsymbol{r}_2) = \varphi(\boldsymbol{r}_1)\varphi(\boldsymbol{r}_2) . \quad (16)$$

The full wavefunction

$$\Psi(\boldsymbol{r}_1, m_{s1}; \boldsymbol{r}_2, m_{s2}) = \psi(\boldsymbol{r}_1, \boldsymbol{r}_2) S(m_{s1}, m_{s2}) \,, \tag{17}$$

where $m_s \left(= \pm \frac{1}{2}\right)$ denotes the z-component of electron spin, is the product of the space part and the spin part. We note that choosing direction $z$ for specifying the component of spin is arbitrary. For the ground-state, the spin part is the antisymmetric function

$$S(m_{s1}, m_{s2}) = \frac{1}{\sqrt{2}} [\alpha(m_{s1})\beta(m_{s2}) - \alpha(m_{s2})\beta(m_{s1})] \,, \tag{18}$$

where $\alpha/\beta$ are the spin wavefunctions with $\alpha\left(\frac{1}{2}\right) = 1$, $\alpha\left(-\frac{1}{2}\right) = 0$, $\beta\left(\frac{1}{2}\right) = 0$, $\beta\left(-\frac{1}{2}\right) = 1$. Note that the full wavefunction is antisymmetric with respect to the exchange of all (including spin) coordinates of two electrons, as it must be because electrons are indistinguishable Fermions.

From now onwards, we will refer to the single-electron wavefunctions $\varphi(\boldsymbol{r})$ as orbitals. The functional form for orbitals in Eq. (16) is taken to be the same as the solution of the single-particle Schrödinger equation corresponding to the respective systems. Thus, for the Hookium

$$\varphi(\boldsymbol{r}) = \left(\frac{a}{\pi}\right)^{3/4} e^{-\frac{ar^2}{2}} \,, \tag{19a}$$

and for the Helium atom

$$\varphi(\boldsymbol{r}) = \left(\frac{a^3}{\pi}\right)^{1/2} e^{-ar} \,, \tag{19b}$$

where $a$ is the variational parameter that is varied to minimize the expectation value of the Hamiltonian. Note that if the electron-electron interaction ($V_{ee}$) was not present, the value of $a$

| Atom/Ion | a [2] | ⟨H⟩ | a | b | ⟨H⟩ | $E_0$[10] |
|---|---|---|---|---|---|---|
| H⁻ | 0.6875 | -0.4727 | 1.01392 | 0.2832 | -0.5133 | -0.5277 |
| He | 1.6875 | -2.8477 | 2.1832 | 1.1885 | -2.8756 | -2.9037 |
| Be²⁺ | 3.6875 | -13.5977 | 4.3872 | 2.9853 | -13.6180 | -13.6555 |
| Ne⁸⁺ | 9.6875 | -93.8477 | 10.7912 | 8.5795 | -93.8476 | -93.9068 |
| Hookium($\omega = \frac{1}{2}$) | 0.4211 | 2.0405 | 0.4211 | 0.4211 | 2.0405 | 2.0000 |

will come out to be equal to $\omega$ for the Hookium and 2 for the Helium atom; these are the exact answers [1,2] for the corresponding non-interacting systems. However, because of $V_{ee}$, electrons will repel each other that makes them less tightly bound. Therefore, when expectation value of the true Hamiltonian is minimized, $a$ should come out to less than $\omega$ (Hookium) and 2 (Helium), as it indeed does.

To calculate the expectation value of the Hamiltonian for each system, one can either do a fully numerical calculation or employ numerical tools after obtaining analytical forms for the expectation value as far as possible. For the wavefunctional form given above, analytical calculation can be performed for Helium and He-like ions. For the Hookium atom, calculations are performed numerically. Results obtained by minimizing ⟨H⟩ with respect to $a$ are shown in **Table 1** in comparison to the exact ones.

**Table 1.** Value of parameters a/(a and b) for wavefunctions of Eq. (19)/Eq. (20) that minimize the expectation value ⟨H⟩ of the Hamiltonian and the corresponding ⟨H⟩. These are compared with the exact values $E_0$ given in the last column. Note that for the Hookium, inclusion of one more parameter does not affect the result. All the numbers are in atomic units.

It is clear from the **Table 1** that variational calculation with even a single parameter leads to reasonable values for the energy. However, the wavefunction itself may not be as accurate as the energy. The reason for this is that in a variational calculation, energy has an accuracy which is one order higher than the accuracy of the wavefunction [2]. Thus, if deviation of a variational wavefunction from the exact one is of $O(\delta)$, the difference in the corresponding energies is of $O(\delta^2)$.

In the wavefunction considered above, both the electrons are in the same orbital. However, for interacting electrons, the orbital extends much more in regions away from the nucleus than the corresponding orbital for the non-interacting system because of the electron-electron interaction. The effect of this interaction on the wavefunction has therefore been taken into account to some extent, but only in the average sense: the potential that the electrons are moving in has been modified by adding to the external potential the electrostatic potential that is given by the electron density. Thus, electrons are assigned to an orbital determined by a mean field. This scheme, where electrons are taken to be in individual orbitals, is therefore known as the ***mean field approximation***. However, in reality something more than this happens. As discussed in the context of the Hookium atom, electrons tend to avoid coming near each other because of their mutual repulsion. As such, when one electron is near the nucleus, the other one will be as far from it as possible while remaining bound to the nucleus so that the total energy is minimized. This makes their motion correlated and this interdependence should be incorporated in the wavefunction, making it a ***correlated wavefunction***. This is certainly not the case if both electrons are in the same orbital resulting from an average potential. Correlation effects can be represented in the wavefunction directly by having terms that are proportional to inter-electronic distance, as in the exact solution (Eq. 13a above) for the Hookium atom. An example for this type of variational wavefunctions is the Hylleraas wavefunction [16] for two-electron atoms and ions forming the He-isoelectronic series.

It is also possible to account for correlations in other ways motivated by physical insight [17]. For example, in the two-electron systems we are discussing, when one electron is near the nucleus, the other one is far from it. Thus, although the orbitals for the two electrons can be taken to have the same exponential form, their coefficient should be different. So, the product wavefunction will be proportional to $e^{-\frac{ar_1^2}{2}}e^{-\frac{br_2^2}{2}}$ ($a \neq b$) for the Hookium and $e^{-ar_1}e^{-br_2}$ ($a \neq b$) for He-like systems, where we now have two variational parameters $a$ and $b$. However, because electrons are indistinguishable, we cannot differentiate between electron 1 and electron 2 and therefore the space part of the wavefunction is made symmetric with respect to exchange of $r_1$ and $r_2$. It is therefore taken to be

$$\psi(\boldsymbol{r}_1, \boldsymbol{r}_2) = C_N \left( e^{-\frac{ar_1^2}{2}} e^{-\frac{br_2^2}{2}} + e^{-\frac{ar_2^2}{2}} e^{-\frac{br_1^2}{2}} \right), \qquad (20a)$$

where $C_N$ is the normalization constant, for the Hookium and

$$\psi(\boldsymbol{r}_1, \boldsymbol{r}_2) = C_N \left( e^{-ar_1} e^{-br_2} + e^{-ar_2} e^{-br_1} \right), \qquad (20b)$$

for He-like systems. We remind the reader again that the full wavefunction is a product of the space part above and the spin part given by Eq. (18). To understand how the wavefunctions in Eqs. (20) keeps two electrons separated, assume $a > b$. Then when one of the electrons is near the center or the nucleus (corresponding to the orbital $e^{-\frac{ar^2}{2}}$ or $e^{-ar}$), the other electron is far from it (being in the orbital $e^{-\frac{br^2}{2}}$ or $e^{-br}$). For He-like systems, expressions for expectation values of different components of $\langle H \rangle$ have been calculated [18] analytically for this wavefunction. For Hookium, calculations are performed numerically. The corresponding values of $\alpha$ and $\beta$ that minimize $\langle H \rangle$ and the corresponding minimum value of $\langle H \rangle$ are given in **Table 1**. Note that for the Hookium, values of $\alpha$ and $\beta$ come out to be the same.

It is clear from the results above that taking a physically motivated parametrized functional form for the wavefunction and optimizing it to minimize $\langle H \rangle$ leads to an upper bound for the ground-state energy. The value of $\langle H \rangle$ approaches the true energy as the number of parameters is increased. Now we ask the question if instead of varying a few parameters in a chosen functional form, the function itself can be varied at each point in space to minimize $\langle H \rangle$ so that the best function describing an orbital is obtained. This indeed is possible and leads to the Hartree and Hartree-Fock methods that we describe next.

**1c. Hartree and Hartree-Fock theories**

We learnt in the section above that with properly chosen approximate wavefunctions, variational method can be applied to estimate the ground-state energy of a system. We now wish to do it systematically within the framework of mean field approximation. This means that we take the many-electron wavefunction to be a product wavefunction made up of single electron orbitals; the best orbitals are then found by minimizing $\langle H \rangle$ calculated with this wavefunction.

In the Hartree method [19, 20], the ground-state wavefunction for $N$ electrons

$$\psi_H(\boldsymbol{r}_1, \boldsymbol{r}_2 \cdots \boldsymbol{r}_i \cdots \boldsymbol{r}_N) = \varphi_1(\boldsymbol{r}_1)\varphi_2(\boldsymbol{r}_2) \cdots \varphi_N(\boldsymbol{r}_N) \qquad (21)$$

is the product of $N$ single-particle orbitals; In writing the ground-state wavefunction in the manner given above, Pauli's principle is taken into account by occupying each orbital by one electron (since 2 electrons can be accommodated in each orbital, there are actually $N/2$ independent orbitals). Thus, the wavefunction in Eq. (21) contains $N$ distinct orbitals with lowest possible energy each. We will refer to this as the Hartree wavefunction.

The expectation value of the Hamiltonian with respect to the Hartree wavefunction, denoted as $E[\varphi_1, \varphi_2, \cdots \varphi_N]$, is

$$E[\varphi_1, \varphi_2, \cdots \varphi_N] = \sum_{i=1}^{N} \left\langle \varphi_i \left| -\frac{1}{2}\nabla^2 \right| \varphi_i \right\rangle + \int \rho(\boldsymbol{r}) \, v_{ext}(\boldsymbol{r}) d\boldsymbol{r} + \frac{1}{2} \iint \frac{\rho(\boldsymbol{r})\rho(\boldsymbol{r}')}{|\boldsymbol{r}-\boldsymbol{r}'|} d\boldsymbol{r} d\boldsymbol{r}'$$

$$-\frac{1}{2}\sum_{i=1}^{N} \iint \frac{|\varphi_i(\boldsymbol{r})|^2 |\varphi_i(\boldsymbol{r}')|^2}{|\boldsymbol{r}-\boldsymbol{r}'|} d\boldsymbol{r} d\boldsymbol{r}' \quad , \qquad (22)$$

From now onwards we will refer to the expectation value of the Hamiltonian as energy also. Here

$$\rho(\boldsymbol{r}) = \sum_{i=1}^{N} |\varphi_i(\boldsymbol{r})|^2 \qquad (23)$$

is the electron density. The notation $E[\varphi_1, \varphi_2, \cdots \varphi_N]$ indicates that the energy is a **functional** of the functions in the square brackets (meaning of a functional and its derivative is explained in **supplemental material**). The first term on the right-hand side of Eq. (22) gives the kinetic energy of electrons and the second term is the energy of interaction with the external potential. The last two terms in the expression arise due to electron-electron interaction. The third term is the Coulomb energy of the electronic charge distribution with charge density $\rho(\boldsymbol{r})$; it is referred to as the **Hartree energy** and is written as $E_H[\rho]$. The last term arises from $i \neq j$ term in the expression for electron-electron interaction energy in the Hamiltonian and is known as the self-interaction energy of electrons. It is understood as follows. In writing the Hartree energy as given above, it is assumed that the associated charge distribution is continuous. However, electrons carry charge -1 (in a.u.) that cannot be divided any further. Therefore, the self-energy of each electron, calculated from its charge density $|\varphi_i(r)|^2$, gets subtracted from the Hartree energy to correct for the granular nature of electronic charge, and leads to the fourth term. We will now apply the variational method to obtain the orbitals that minimize the energy.

> Assume $N$ electrons distributed uniformly inside a sphere of radius $R$ atomic units. Calculate their Hartree energy and the total self-energy.

Suppose the energy attains its minimum value for the set $\{\varphi_1(r_1), \varphi_2(r_2) \cdots \varphi_N(r_N)\}$. If orbitals are now varied arbitrarily about these to $\{\varphi_1(r_1) + \delta\varphi_1(r_1), \varphi_2(r_2) + \delta\varphi_2(r_2) \cdots \varphi_N(r_N) + \delta\varphi_N(r_N)\}$, while keeping each of them normalized up to the first order so that

$$\int \delta\varphi_i^*(r)\varphi_i(r)dr + \int \varphi_i^*(r)\delta\varphi_i(r)\,dr = 0 \tag{24}$$

for each $i$, the corresponding change $\delta E$ in the energy calculated up to the first order in $\{\delta\varphi_1(r_1), \delta\varphi_2(r_2) \cdots \delta\varphi_N(r_N)\}$ will vanish. This change is given as

$$\delta E = \sum_{i=1}^{N}\left[\int \delta\varphi_i^*(r)\left(-\frac{1}{2}\nabla^2 + v_{ext}(r) + \int \frac{\rho(r')}{|r-r'|}dr' - \int \frac{|\varphi_i(r')|^2}{|r-r'|}dr'\right)\varphi_i(r)\,dr \right.$$
$$\left. + \int \varphi_i^*(r)\left(-\frac{1}{2}\nabla^2 + v_{ext}(r) + \int \frac{\rho(r')}{|r-r'|}dr' - \int \frac{|\varphi_i(r')|^2}{|r-r'|}dr'\right)\delta\varphi_i(r)\,dr\right] \tag{25}$$

Since each of the variations $\{\delta\varphi_1(r_1), \delta\varphi_2(r_2) \cdots \delta\varphi_N(r_N)\}$ is arbitrary and satisfies the normalization condition given by Eq. (24), $\delta E$ will vanish if

$$\left(-\frac{1}{2}\nabla^2 + v_{ext}(r) + \int \frac{\rho(r')}{|r-r'|}dr' - \int \frac{|\varphi_i(r')|^2}{|r-r'|}dr'\right)\varphi_i(r) = \epsilon_i\varphi_i(r)\;, \tag{26}$$

where $\epsilon_i$ is a constant and is different for each orbital. It is left as an exercise for the reader to show that this is the case using the fact that both $\varphi_i(r)$ and $\varphi_i^*(r)$ satisfy Eq. (26) and vanish as $|r| \to \infty$ for bound states. We direct the reader to go to **supplemental material** to learn how this derivation is done using **functional derivatives**; In what follows, we will make use of it directly to minimize different functionals.

Let us now understand Eq. (26) physically. To do this, we look at the effective potential

$$v_{ext}(r) + \int \frac{\rho(r')}{|r-r'|}dr' - \int \frac{|\varphi_i(r')|^2}{|r-r'|}dr' \tag{27}$$

seen by the electron in orbital $\varphi_i(r)$. The potential is the sum of the external potential and the Coulomb potential of the rest of the electrons. The latter is obtained by subtracting the **self-interaction potential** (third term in the equation above) from the Coulomb potential (second term in the equation and known as the **Hartree potential**) of the total electronic charge density $\rho(r)$. Thus, in Hartree theory the effect of electron-electron interaction is accounted for taking the electrons to be moving in a **mean field** given by Eq. (27), i.e., electrons are not moving in a correlated manner. This is reflected in the wavefunction with the absence of terms dependent on distance between two electrons, and in the potential seen by the electrons being the average potential. This potential is orbital-dependent because of self-interaction of an electron depends on its orbital.

Equation (26) is known as the **Hartree equation**. Since the effective potential in it depends on the orbitals themselves, the equation is solved **self-consistently**. Let us understand what that means. Suppose we choose a set $\{\varphi_1(\boldsymbol{r}_1), \varphi_2(\boldsymbol{r}_2) \cdots \varphi_N(\boldsymbol{r}_N)\}$ of orbitals and use them as input to calculate the effective potential. When the Hartree equation is solved with this potential, the resulting output orbitals will not necessarily be the same as the ones we started with. Therefore, the new set of orbitals is again used to construct the effective potential and Hartree equation is solved again to get the next set of orbitals and this process is repeated until the input and output orbitals are the same. These orbitals represent the **self-consistent** solution of the Hartree equation. As is clear, obtaining self-consistent solution is an iterative process. A little reflection on the procedure of obtaining self-consistent solution clearly indicates that this is to be done numerically using a computer. An algebraic example of self-consistent solution is given in the box below.

### *An algebraic example of a self-consistent calculation*

Consider the quadratic equation

$$x^2 + x - 6 = 0 \quad ,$$

Which has the solution $x = -3$ and $x = 2$. We now will get these solutions by starting with a guess for the solution and making it self-consistent iteratively. For this we write the equation above as

$$x^2 = 6 - x$$

and connect the solution in the $i^{th}$ iteration to that in the $(i + 1)^{th}$ iteration according to

$$x_{i+1} = \pm\sqrt{6 - x_i}$$

We start with a guess of $x_0 = 1$. Then the series of answers one gets by keeping positive and negative solutions are, respectively,

$$1, 2.236, 1.940, 2.015, 1.996, 2.001$$

and

$$1, -2.236, -2.870, -2.978, -2.996$$

As one can see, the solutions are slowly converging towards the correct values. One the other hand, one could have also rearranged the equation differently and do the iteration according to

$$x_{i+1} = 6 - x_i^2$$

Again, if we start with $x_0 = 1$, we find that the solution starts becoming larger and larger and does not converge. In other words, the process is unstable. To make it stable, the new input value of $x$ is taken to be a weighted mixture of $x_i$ and $x_{i+1}$. In the present case, we will take it to be

$$0.8 x_i + 0.2 x_{i+1}$$

Now starting with $x_0 = 1$, we get a series of solutions as follows.

| $x_i$ | $x_{i+1}$ | $x_{input} = 0.8x_i + 0.2x_{i+1}$ | $x_{output} = 6 - x_i^2$ |
|---|---|---|---|
| 1 | 5 | 1.8 | 2.76 |
| 1.8 | 2.76 | 1.992 | 2.032 |
| 1.992 | 2.032 | 2.000 | 2.000 |

On the other hand, the solution $x = -3$ is not obtained using the formula $x_{i+1} = 6 - x_i^2$ iteratively. So, the process remains unstable for this solution. This example shows that in general, an iterative process is stable if the new input is taken to be a judicious mixture of the old input and the output from it. Thus, performing a self-consistent calculation is as much an art as it is a technical skill.

**Exercise:** Write the Hartree equation for He-like two-electron systems. Connection between this equation and the variational calculation done earlier can be seen as follows. Calculate the potential for the Hartree equation using orbital of Eq. (19b). Then take the trial wavefunction to be of the same form as that of Eq. (19b) with $a$ replaced by a different parameter $b$ and calculate the expectation value of the Hamiltonian of the Hartree equation. Next, minimize this expectation value with respect to $\beta$, keeping $\alpha$ fixed. After this, take $a = b$ in the resulting equation to make them self-consistent and solve for their value. This will give $a = Z - \frac{5}{16}$.

One system for which exact analytical solution of Hartree equation exists is the homogeneous electron gas (HEG). HEG is a collection of electrons in the background of equal amount of positive charge spread uniformly with constant density $\rho$. Explicit solutions for the HEG plays an important role in the development of energy functionals in density-functional theory. Hence, we discuss it below in detail.

For free electrons (constant external potential and no interaction potential energy), it is well known that the solutions of the Schrödinger equation are plane waves. When normalized over a large volume $V$ with periodic boundary conditions, these are given as $\sqrt{\frac{1}{V}} e^{i\mathbf{k}\cdot\mathbf{r}}$ and various properties of free electron gas are obtained [15, 22] using them. Self-consistent solutions of Hartree equation too are plane waves. This can be easily seen as follows. Plane waves give rise to a uniform density of electrons equal to the background density. Furthermore, the self-interaction potential for each orbital is zero for an infinite system. Thus, the net effective potential of Eq. (27) isa constant. This then gives plane waves as the self-consistent solutions of Hartree equation. In calculating the energy also, it is easily seen that the energy of interaction between the electrons and the uniform positive background cancels with the sum of electrostatic energy of the background charge and electronic charge densities. The energy of HEG in Hartree theory is therefore the same as that for free electrons and equal to their kinetic energy. For density $\rho$, its value per electron is (in atomic units)

$$\epsilon_k(\rho) = \frac{3}{10}(3\pi^2\rho)^{\frac{2}{3}} \quad , \tag{28a}$$

which is equal to

$$\frac{3}{5}\left(\frac{k_F^2}{2}\right) \quad , \tag{28b}$$

where $k_F = (3\pi^2\rho)^{\frac{1}{3}}$ is the Fermi wavevector, i.e. the wavevector for the highest occupied quantum-state. Thus, kinetic energy per electron for a HEG is three-fifths of the energy of the highest occupied state. It is also expressed in terms of **Wigner-Seitz radius** $r_s$ as

$$\epsilon_k(r_s) = \frac{1.105}{r_s^2} \quad , \tag{28c}$$

where $r_s$ is defined by the equation

$$\rho = \left(\frac{4\pi}{3}r_s^3\right)^{-1} \quad , \tag{29}$$

and gives the radius of the sphere containing one electron.

The question we now ask if the Hartree wavefunction is the best possible product wavefunction or can it be improved further? The answer is that the wavefunction can be made better. It is done by considering all possible ways of distributing $N$ electrons among $N$ orbitals and then taking linear combination of these products so that the final wavefunction is antisymmetric with respect to exchange of the coordinates (including spin) of any two electrons. The best orbitals are then found by applying the variational principle. The wavefunction thus obtained is known as the Hartree-Fock (HF) wavefunction, and the method as Hartree-Fock theory [3,15,23,24]. It gives the best possible mean-field wavefunction and energy for a system. We now discuss the theory in detail.

The wavefunction in HF theory is constructed from spin-orbitals

$$\chi_{i,s}(x) = \varphi_{i,s}(\mathbf{r})s(m_s) \ , \tag{30}$$

where $x = (\mathbf{r}, m_s)$ is the coordinate of an electron including the spin-variable $m_s$ that gives the z-component of its spin and takes values $\pm\frac{1}{2}$. The spin wavefunction $s(m_s)$ could be either $\alpha(m_s)$ or $\beta(m_s)$. As indicated earlier, the values taken by the spin wavefunction are

$$\alpha\left(+\tfrac{1}{2}\right) = 1, \quad \beta\left(-\tfrac{1}{2}\right) = 0, \quad \alpha\left(+\tfrac{1}{2}\right) = 0, \quad \beta\left(-\tfrac{1}{2}\right) = 1 \ , \tag{31a}$$

It is evident that products of two spin wavefunctions for different combinations of spin quantum number $s$ and spin variable $m_s$ are

$$\alpha^*(m_s)\alpha(m_s') = \delta_{m_s,\frac{1}{2}}\delta_{m_s,m_s'} \ ,$$

$$\beta^*(m_s)\beta(m_s') = \delta_{m_s,-\frac{1}{2}}\delta_{m_s,m_s'} \ ,$$

$$\alpha^*(m_s)\beta(m_s) = 0 \ , \tag{32}$$

The integration $\int dx$ with respect to $x$ is then equivalent to $\int d\mathbf{r} \int dm_s$ with the symbolic integration with respect to $m_s$ being the summation over its discreet values, i.e.,

$$\int f(m_s)\, dm_s \equiv \sum_{m_s} f(m_s) = f\left(+\tfrac{1}{2}\right) + f\left(-\tfrac{1}{2}\right) \ . \tag{33}$$

Using Eqs. (32) and (33), it is clear that spin wavefunctions are normalized. Finally, by including index $s$ in the orbital $\varphi_{i,s}(\mathbf{r})$, we have shown explicitly that besides other quantum numbers denoted by $i$, it may have a dependence on $s$.

In terms of spin orbitals, the HF wavefunction for $N$ electrons is the determinant

$$\psi_{HF}(x_1, x_2 \cdots x_N) = \sqrt{\frac{1}{N!}} \begin{vmatrix} \chi_1(x_1) & \chi_1(x_2) & \cdots & \chi_1(x_N) \\ \chi_2(x_1) & \chi_2(x_2) & \cdots & \chi_2(x_N) \\ \vdots & \vdots & & \vdots \\ \vdots & \vdots & & \vdots \\ \chi_N(x_1) & \chi_N(x_2) & \cdots & \chi_N(x_N) \end{vmatrix} \ , \tag{34}$$

and is known as the Slater determinant. Like the product wavefunction of Eq. (21), in HF wavefunction also each spin orbital is occupied by one electron. However, unlike the wavefunction of Eq. (21), HF wavefunction does not assign a specific electron, identified by the subscript on coordinate $x$, to a given spin orbital. This is reflected in the wavefunction being a linear combination of products of orbitals with each product having a different permutations of electron labels distributed among the orbitals. Therefore, quantum numbers associated with a spin orbital in the determinant above cannot be assigned to a particular electron, consistent with the indistinguishability of electrons.

Next, we take the expectation value of the Hamiltonian with respect to the HF wavefunction. For this we make use of the following relation [23, 24]:

$$\langle\psi_{HF}|\hat{O}|\psi_{HF}\rangle = \langle \chi_1(x_1)\chi_2(x_2)\cdots\chi_N(x_N)|\hat{O}|\begin{bmatrix} \chi_1(x_1) & \chi_1(x_2) & \cdots & \chi_1(x_N) \\ \chi_2(x_1) & \chi_2(x_2) & \cdots & \chi_2(x_N) \\ \vdots & \vdots & & \vdots \\ \vdots & \vdots & & \vdots \\ \chi_N(x_1) & \chi_N(x_2) & \cdots & \chi_N(x_N) \end{bmatrix}\rangle \; , \quad (35)$$

where in general the operator $\hat{O} = \hat{O}(x_1, x_2 \cdots x_N)$ may also depend on spin of electrons. Therefore, the expectation value is taken by integrating over the spatial as well as spin coordinates of the electrons. Notice that the factors involving $N!$ do not appear in the expression above. Of our interest right now are the one-particle operators and two-particle operators. One-particle operators are written as

$$\hat{O}_1(x_1, x_2 \cdots x_N) = \sum_{i=1}^{N} \hat{O}_1(r_i) \; . \quad (36)$$

and include the density, the kinetic energy and the external energy operators. Expectation value of these operators is given, using Eq. (35), as [23, 24]

$$\langle\psi_{HF}|\hat{O}_1|\psi_{HF}\rangle = \sum_{i=1}^{N} \int \chi_i^*(x_i)\hat{O}_1(x_i)\chi_i(x_i)dx_i \; . \quad (37a)$$

Since the integration variable can be written by any symbol, we write the equation above as

$$\langle\psi_{HF}|\hat{O}_1|\psi_{HF}\rangle = \sum_{i=1}^{N} \int \chi_i^*(x)\hat{O}_1(x)\chi_i(x)dx \; . \quad (37b)$$

We now use the explicit form of the spin orbitals and separate the spin and other quantum numbers to write the expression in Eq. (37b) as

$$\langle\psi_{HF}|\hat{O}_1|\psi_{HF}\rangle = \sum_{s=\alpha}^{\beta} \int s^*(m_s)s(m_s)dm_s \sum_i \int \varphi_{i,s}^*(r)\hat{O}_1(r)\varphi_{i,s}(r)dr \; . \quad (37c)$$

Finally, using Eq. (32), or equivalently the normalization condition for the spin wavefunctions, the expectation value of the single-particle operator of Eq. (36) is

$$\langle\psi_{HF}|\hat{O}_1|\psi_{HF}\rangle = \sum_{m_s} \sum_i \int \varphi_{i,m_s}^*(r)\hat{O}_1(r)\varphi_{i,m_s}(r) \, dr \; . \quad (38)$$

(Note that we have replaced $s$ by $m_s$ in the labels for the orbitals employing Eq. (32)). For example, using Eq. (38), electron density is given as

$$\rho(r) = \sum_{m_s} \sum_i \varphi_{i,m_s}^*(r)\varphi_{i,m_s}(r) \; . \quad (39)$$

We now consider two-particle operators. These involve coordinates of two electrons and have the following form

$$\hat{O}_2(x_1, x_2 \cdots x_N) = \frac{1}{2}\sum_{\substack{i,j=1 \\ (i \neq j)}}^{N} \hat{O}_2(r_i, r_j) \; . \quad (40)$$

Using Eq. (35) and following the steps involved in obtaining the expectation value for single-particle operators, we get [23, 24]

$$\langle\psi_{HF}|\hat{O}_2|\psi_{HF}\rangle = \frac{1}{2}\sum_{m_s,m_s'} \sum_{i,j} \left[ \int\int \varphi_{i,m_s}^*(r)\varphi_{j,m_s'}^*(r')\hat{O}_2(r,r')\varphi_{i,m_s}(r)\varphi_{j,m_s'}(r')dr\,dr' \right.$$

$$-\frac{1}{2}\sum_{m_s,m_s'}\sum_{i,j}\left[\delta_{m_s m_s'}\iint \varphi_{i,m_s}^*(\bm{r})\varphi_{j,m_s'}^*(\bm{r}')\hat{O}_2(\bm{r},\bm{r}')\varphi_{i,m_s'}(\bm{r}')\varphi_{j,m_s}(\bm{r})d\bm{r}\,d\bm{r}'\right]. \quad (41)$$

Notice that now the sum over orbital quantum numbers does not have $(i, m_s) \neq (j, m_s')$ terms. This condition is automatically satisfied since $(i, m_s) = (j, m_s')$ terms in the two expressions on the right-hand side of Eq. (41) cancel.

The two-particle operator of our interest right now is the electron-electron interaction energy term with

$$\hat{O}_2(\bm{r}_i,\bm{r}_j) = \frac{1}{|r_i - r_j|}. \quad (42)$$

Using the formula for the density given in (39) and the explicit form of Coulomb potential, the first expression in Eq. (41) is easily shown to be the Hartree energy, which was also a component of the energy in Hartree theory. We call this the direct term. The second expression, known as the exchange term, gives an additional component of electron-electron interaction energy and is referred to as the **exchange energy**. This arises from that product in the determinant where the coordinates of electrons in orbitals $\chi_{i,s}(x)$ and $\chi_{j,s'}(x')$ have been swapped in the diagonal term while all the other coordinates remain the same. This term therefore picks up a minus sign with respect to the direct term and these orbitals now appear as $\chi_{i,s}(x')$ and $\chi_{j,s'}(x)$. Because of this, integration over spin coordinates gives the Kronecker delta $\delta_{m_s m_s'}$ in the expression for the exchange energy. What this means is that contribution to exchange energy comes only from interaction between electrons in orbitals with the same spin, or z-component of spin to be more precise. When the sum over $m_s'$ is carried out and $\hat{O}_2(\bm{r},\bm{r}')$ is replaced by $\frac{1}{|r-r'|}$, explicit expression for the exchange energy $E_x$ comes out to be

$$E_x = -\frac{1}{2}\sum_{m_s}\sum_{i,j}\left[\iint \frac{\varphi_{i,m_s}^*(\bm{r})\varphi_{j,m_s}^*(\bm{r}')\varphi_{i,m_s}(\bm{r}')\varphi_{j,m_s}(\bm{r})}{|r-r'|}d\bm{r}d\bm{r}'\right], \quad (43)$$

where the sum in the expression above is over the occupied orbitals. Note that the spin quantum number of all the orbitals in the expression above is the same. In addition, as right after Eq. (41), self-energy of electron in each occupied orbital is included in the exchange energy.

Collecting all the terms together, the expression for energy in HF theory is

$$E_{HF} = \sum_{m_s}\sum_i \int \varphi_{i,m_s}^*(\bm{r})\left(-\frac{1}{2}\nabla^2\right)\varphi_{i,m_s}(\bm{r})\,d\bm{r} + \int \rho(\bm{r})v_{ext}(\bm{r})d\bm{r} + \frac{1}{2}\iint \frac{\rho(\bm{r})\rho(\bm{r}')}{|r-r'|}d\bm{r}d\bm{r}'$$

$$-\frac{1}{2}\sum_{m_s}\sum_{i,j}\left[\iint \frac{\varphi_{i,m_s}^*(\bm{r})\varphi_{j,m_s}^*(\bm{r}')\varphi_{i,m_s}(\bm{r}')\varphi_{j,m_s}(\bm{r})}{|r-r'|}d\bm{r}d\bm{r}'\right]. \quad (44)$$

Taking the functional derivative (see **supplemental material**) of $E_{HF}$ with respect to complex conjugate of an orbital and setting it to zero with the constraint that each orbital is normalized gives the Hartree-Fock equation

$$\left(-\frac{1}{2}\nabla^2 + v_{ext}(\bm{r}) + \int \frac{\rho(\bm{r}')}{|r-r'|}d\bm{r}'\right)\varphi_{i,m_s}(\bm{r}) - \sum_j \int \frac{\varphi_{j,m_s}^*(\bm{r}')\varphi_{i,m_s}(\bm{r}')\varphi_{j,m_s}(\bm{r})}{|r-r'|}d\bm{r}'$$

$$= \epsilon_{i,m_s}\varphi_{i,m_s}(\bm{r}), \quad (45a)$$

for that orbital. As is the case with the Hartree equation, HF equation also is solved self-consistently since the potential depends on the solution itself. Substitution of the solution of HF equation in Eq. (44) gives the HF energy.

The effective potential in HF theory consists of two parts: the Hartree potential and the exchange potential. The exchange potential is a non-local potential in that it acts on the orbital as an integral operator. Furthermore, it also depends on the spin of the orbital that it is acting upon.

> **Comment:** General operation of a non-local potential $v(\boldsymbol{r},\boldsymbol{r}')$ on a function $f(\boldsymbol{r})$ is given as
>
> $$\int v(\boldsymbol{r},\boldsymbol{r}')f(\boldsymbol{r}')d\boldsymbol{r}' \ .$$
>
> For multiplicative potentials, like $v_{ext}(\boldsymbol{r})$ or the Hartree potential, $v(\boldsymbol{r},\boldsymbol{r}') \propto \delta(\boldsymbol{r}-\boldsymbol{r}')$.

We write the exchange potential as

$$v_{x,m_s}(\boldsymbol{r},\boldsymbol{r}') = -\sum_j \frac{\varphi^*_{j,m_s}(\boldsymbol{r}')\varphi_{j,m_s}(\boldsymbol{r})}{|\boldsymbol{r}-\boldsymbol{r}'|} \ , \tag{46}$$

where, as noted earlier, the sum in the expression above is over the occupied orbitals. It is evident that the exchange potential includes in it the self-interaction potential of Hartree equation and additional terms, making it more general. The additional terms give the interaction between electrons in different orbitals. Using expression of Eq. (46), the Hartree-Fock equation is rewritten as

$$\left(-\frac{1}{2}\nabla^2 + v_{ext}(\boldsymbol{r}) + \int \frac{\rho(\boldsymbol{r}')}{|\boldsymbol{r}-\boldsymbol{r}'|}d\boldsymbol{r}'\right)\varphi_{i,m_s}(\boldsymbol{r}) + \int v_{x,m_s}(\boldsymbol{r},\boldsymbol{r}')\varphi_{i,m_s}(\boldsymbol{r}')d\boldsymbol{r}'$$

$$= \epsilon_{i,m_s}\varphi_{i,m_s}(\boldsymbol{r}) \quad (45b)$$

For a general discussion on the physical interpretation of the exchange energy and potential and how the exchange potential can be made local on the basis of this interpretation, we refer the reader to reference [25].

Some general features of HF theory are listed below:

(i) HF theory is the best mean-field theory as no other product wavefunction can give a lower energy than the HF energy;
(ii) Negative of the eigenvalues $\epsilon_{i,m_s}$ of HF equation can be shown to approximate the removal energy of an electron from the corresponding orbital. This is known as **Koopmans' theorem** [24,26];
(iii) For a single electron system, the direct and the exchange terms cancel in both the total energy and in the effective potential. This clearly indicates that the theory is free of self-interaction of electrons;
(iv) For homogeneous electron gas (HEG), self-consistent solutions of HF equation also are plane waves. This is so because, like in Hartree theory, when these solutions are substituted in the expression for the potential, all the terms in it come out to be constants. However, the energy now has the additional component of exchange energy. Its value per electron in terms of the density is [15]

$$\epsilon_x(\rho) = -\frac{1}{4}\left(\frac{3\rho}{\pi}\right)^{\frac{1}{3}} \quad , \qquad (47a)$$

or equivalently

$$\epsilon_x(r_s) = -\frac{0.456}{r_s} \qquad (47b)$$

in terms of the Wigner-Seitz radius $r_s$;

(v) The total average energy per electron in Hartree-Fock theory is [15]

$$\epsilon(r_s) = \frac{1.105}{r_s^2} - \frac{0.456}{r_s} \quad . \qquad (48)$$

Next, we present some representative results obtained in Hartree-Fock theory and compare them with experimental values. We start with the results for atoms. Hartree-Fock equation for atoms has been solved [27, 28] fully numerically and also by expanding the orbitals in terms of basis functions known as Slater orbitals. The HF orbitals obtained using the latter method have been tabulated [29]. These semi-analytical orbitals are quite useful when we wish to perform any calculations for atoms by employing HF wavefunctions.

**Exercise:** Write a computer program to calculate for atoms the total energy and its different components employing their HF orbitals given in ref. [29].

Given in **Table 2** are the total energies and negative $-\epsilon_{max}$ of the energy eigenvalues for the highest occupied orbital (HO) in HF theory [29] for the hydrogen anion and some noble gas atoms along with the experimental [30,31] total energies and ionization potential. It is seen that the Hartree-Fock energies always lie above the experimental values. The difference arises because HF wavefunction is not the exact wavefunction. Since this wavefunction is calculated from an average potential, it does not have terms dependent on interelectronic distances and therefore neglects correlation between electrons arising due to their Coulomb interaction. Inclusion of these effects will lower the energy and bring it closer to the true ground-state energy. Furthermore, for heavy atoms, relativistic effects also contribute the difference. We define the correlation energy $E_c$ of a system as

$$E_c = E_{HF} - E_{exact}^{non-relativistic} \quad . \qquad (49)$$

It follows from the definition that correlation energies will always be negative. Since relativistic effects are negligible for atoms shown in **Table 2**, their correlations energies can be calculated from the numbers given in the Table.

**Table 2.** Total energy and eigenenergy of the highest occupied (HO) orbital of H⁻ and three noble gas atoms as calculated in HF theory. These are compared with the experimental total energies and ionization potentials. All numbers are given in atomic units.

| Atom | Total energy | | Eigenvalue (HO) and Ionization potential | |
|---|---|---|---|---|
| | E<sub>HF</sub> [29] | E<sub>expt</sub>[30, 31] | $-\epsilon_{max}$ [29] | I<sub>expt</sub> [30, 31] |
| H⁻ | -0.4879 | -0.5277 | 0.0462 | 0.0277 |
| He | -2.8617 | -2.9034 | 0.9179 | 0.9035 |
| Ne | -128.5471 | -129.0600 | 0.8504 | 0.7925 |
| Ar | -526.8174 | -529.2490 | 0.5910 | 0.5790 |

> **Exercise:** Calculate what percent of the total energy is the correlation energy in the atomic systems shown in **Table 2**? Whys does this percentage become smaller with increasing atomic number?

A look at the numbers in **Table 2** indicates that HF method is a good approximation for calculating total energies and also gives reasonable estimates of the ionization energies. Why should we then be concerned with correlation energy? The answer is provided by the value of energy for $H^-$ and when we compare it with the energy of hydrogen atom, which is $-0.5$ atomic units. Since in HF theory the energy of $H^-$ is above the energy of hydrogen atom, the anion should spontaneously release one electron and become neutral so that the total energy is lowered, if the HF energy were its true energy. However, correlation makes the true energy of $H^-$ lower than that of hydrogen atom. Although the correlation energy is numerically small here, it plays a key role in making the ion stable. This is only one of many examples where the character of a system will be totally different if correlation is not taken into account.

Now we present the example of extended systems when HF theory is applied to them. We first discuss the results for HEG. HEG is best suited to metals where the ionic potential is screened by the electrons and the resulting background potential can therefore be approximated as uniform. Energy of HEG in HF theory has already been given. Here our focus is on results for some properties other than the energy. These are not given [15] correctly by HF theory for metals. First the bandwidth of the conduction band comes out to be much larger than seen in experiments. Secondly, the density of states for electrons becomes infinitely large at the Fermi level and that leads to wrong temperature dependence of specific heat at low temperatures.

Let us next discuss what are the results when HF theory is applied to other systems, with the potential of the ions replacing the constant background potential of HEG. In **Table 3**, we present the results for lattice constants, bulk moduli and energy band gaps of some non-metallic systems.

**Table 3.** Lattice constants, bulk modulus and energy band gap for some non-metallic systems calculated in HF theory. These are compared with the corresponding experimental numbers.

| Solid | $a_{HF}$ [Å] | $a_{Expt}$ [Å] | $B_{HF}$ [GPa] | $B_{Expt}$ [GPa] | $E_{Gap}^{HF}$ [eV] | $E_{Gap}^{Expt}$ [eV] |
|---|---|---|---|---|---|---|
| C | 3.55 [32] 3.57 [33] | 3.57 [22] | 438 [32] 464 [33] 476 [33] | 442 [34] | 12.1 [35,36] | 5.5 [22] |
| Si | 5.46 [33] | 5.43 [22] | 109 [33] | 99 [34] | 5.6 [36] | 1.2 [22] |
| Ge | 5.79 [33] | 5.65 [22] | 85 [33] | 76 [34] | 4.2 [36] | 0.7 [22] |
| Ne | 4.43 | 4.46[37] | 1.2 | 1.1[38] | 25.4 [39] 25.2[40] | 21.4 [41] |
| Ar | 5.15 | 5.30[42] | 4.4 | 2.9[38] | 18.5 [39] 18.5[40] | 14.3 [41] |
| Kr | 5.50 | 5.65[43] | 5.1 | 3.3[38] | 16.4[40] | 11.6[41] |
| Xe | 6.01 | 6.13[44] | 5.4 | 3.7[38] | --- | 9.8 [41] |

It is clear from the **Table 3** that lattice constants of the materials considered calculated with HF theory are close to the experimental results. However, the bulk modulus is reasonable for some of these and is overestimated for others. On the other hand, the band gaps in HF theory are consistently larger than the experimental band-gaps for all systems. Overestimation of band gaps is easily understood when we look at Eq. (45b) for the unoccupied orbital $\varphi_C(r)$ at the bottom of the conduction band and the operation of the exchange potential of Eq. (46) on this orbital. It is first noted that in calculating the exchange potential, the sum is over only the occupied orbitals and the Hartree potential is calculated from the density of electrons occupying these orbitals. However, we can add the density $|\varphi_C(r')|^2$ of the unoccupied orbital to $\rho(r')$ and extend the sum in the exchange term to include $\varphi_C(r)$ without affecting the equation. But looking at the equation in this manner shows the HF equation for $\varphi_C(r)$ to be for a system that has an additional electron. This makes the resulting sum of the Hartree and exchange potential more positive for $\varphi_C(r)$ and gives the orbital energy for it also to be more positive, corresponding to the system with an additional electron. This gives a larger gap between the bottom of the conduction band and top of the valence band than expected for the system with a fixed number of electrons.

**1d. Going beyond Hartree-Fock theory**

As must be clear by the presentation above, HF theory is a good first approximation to the true many-body solution but also fails considerably in predicting certain properties. This happens because HF wavefunction neglects Coulomb correlations among electrons. The example we saw above in this connection is that of $H^-$ ion. In such two electron systems, because of Coulomb correlations among the electrons, the wavefunction has in it a term dependent on $r_{12}$. This makes the charge cloud of one electron shift away from the nucleus resulting in bringing the nucleus closer to the second electron and leading to an attraction between them. That makes it possible for the second electron to get bound to the system. Without such a correlation between the electrons, it will not be possible to bind the extra electron to a neutral hydrogen atom rendering the $H^-$ ion to be unstable. This is shown schematically in **Figure 1**.

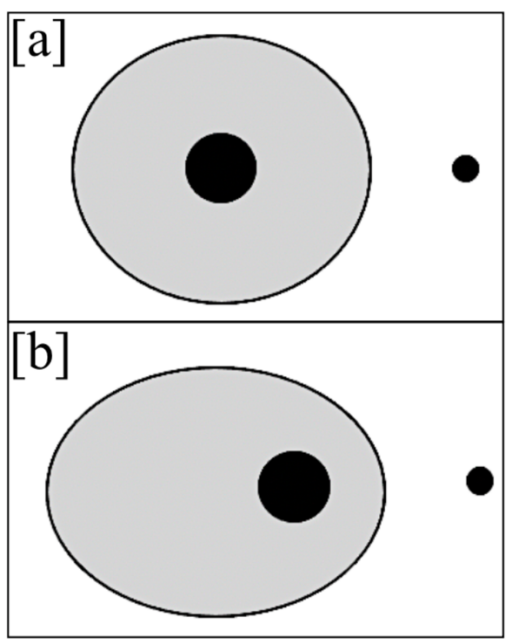

**Figure 1.** (a) A neutral hydrogen atom shown schematically by a nucleus, depicted by the large dark circle, and a spherically symmetric electron charge cloud around it. The picture shows an imaginary situation where the charge cloud of the electron in the atom remains unaffected as another electron, shown by small dark circle, approaches it. In such a case there is no attraction between the atom and the approaching electron. This happens if correlation between the electrons is neglected; (b) In contrast to the situation in (a), the charge cloud around the nucleus is distorted because of electron-electron interaction showing that the correlation between them has been taken into account. Because of the resulting asymmetric charge distribution in the atom the second electron is attracted to it.

In metals, the long-range component of Coulomb interaction among the electrons gives rise to collective motion of electrons. The residual interaction between individual electrons is thus reduced to being short-range and, therefore, weak. This implies that electrons can be treated as effectively free, which removes the problems arising in HF theory. Separating the motion of electrons into collective and individual coordinates and dealing with these differently is then one way [3, 45] of taking care of correlation effects in metals. This method of obtaining the correlation energy has been applied making an approximation known as the random phase approximation [45], or simply the RPA. An equivalent way of calculating the correlation energy within the RPA is by integrating the frequency-dependent response function of HEG [46,47].

Notwithstanding the discussion in the paragraphs above, the question arises as to how does one go beyond Hartree-Fock theory and account for Coulomb correlations in the wavefunction in general? We discuss this qualitatively in the following paragraphs without doing any derivations or performing any rigorous calculations. The idea is to convey that wavefunctional calculations become complicated when one goes beyond mean-field theories. This suggests need to develop alternatives to wavefunctional theory, and density-functional theory provides precisely that.

One approach to account for Coulomb correlations in a wavefunction is to explicitly incorporate in it terms that depend on distance between electrons. Hyllerass wavefunctions [16] for two-electron atomic systems, alluded to in section (1b), are an example of this. These wavefunctions have explicit dependence on $r_{12}$ and give highly accurate energies for two-electron systems. Other correlated wavefunctions in section (1b) were given by Eqs. (20a) and (20b). These were constructed in such a way that when one electron was near the nucleus, the other one stayed far from it. However, when number of electrons becomes more than two, it is not practically possible to include in the wavefunction terms dependent on all inter-electronic distances nor is it easy to construct functions like those given in Eqs. (20a) and (20b). How does one then make further progress?

An obvious path to go beyond Hartree-Fock theory is to expand the wavefunction in terms of Slater determinants constructed from the ground- and excited-state orbitals. The first term in such an expansion is the Hartree-Fock wavefunction. The wavefunction so constructed therefore involves many more orbitals than the number of electrons and also a large number of expansion parameters. This makes evaluation of various expectation values and optimization of these parameters difficult and numerically cumbersome. The methods based on this approach are the multi-configuration Hartree-Fock (MCHF) and configuration-interaction (CI) methods [23]. Readers are requested to go to the literature to learn more about these.

The model Hamiltonian approach is another way to account for the Coulomb correlations. The Pariser-Parr-Pople [48-50] or equivalently the Hubbard Hamiltonian [51] is one such model Hamiltonian. The former was proposed to deal with electron-electron correlations in unsaturated molecules and the latter with electrons in solids. These have been extended further to develop model Hamiltonians according to the problem to be addressed.

As a demonstration of the calculation of correlation energy, we take the example of Wigner crystal [52,53] where it can be calculated analytically. Wigner observed that for extremely low density ($r_s \to \infty$) HEG, electrons settle to form a body-centered-cubic (BCC) crystal, known as the Wigner crystal. Using this information, we can calculate the correlation energy per electron for HEG in the limit of $r_s \to \infty$ as follows. We consider the Wigner crystal made up of neutral spheres of radius $r_s$ with an electron at the centre of the sphere and the background positive charge spread uniformly over the sphere. Since each sphere is neutral, there is no energy of interaction between the spheres. Therefore, the total energy per electron in this system will be the sum of the interaction-energy between the electron and the positively charged sphere and the self-energy of positive charge sphere. It is equal to

$$-\frac{3}{2}\left(\frac{1}{r_s}\right) + \frac{3}{5}\left(\frac{1}{r_s}\right) = -\frac{0.9}{r_s} \quad . \tag{50}$$

**Exercise:** Check from reference [54] that the relative difference between the ground state energy for paramagnetic state of HEG and the value given by Eq. (50) becomes smaller with increasing $r_s$.

Using Eqs. (48) and (49), we get the correlation energy per electron (neglecting the kinetic energy term $1.105/r_s^2$ as $r_s \to \infty$) for very low density HEG with equal number of electrons of both spins as

$$\epsilon_c(r_s) = -\frac{0.444}{r_s} \quad . \quad (51a)$$

Based on this, Wigner interpolation formula [55], which is an approximate formula valid for a range of spin-compensated electron densities is given as

$$\epsilon_c(r_s) = -\frac{0.440}{r_s + 7.85} \quad . \quad (51b)$$

With this we conclude the introduction to wavefunctional methods. We have pointed out the difficulties that finding solutions of many-electrons Schrödinger equation poses. In particular, since the wavefunction is a function of $3N$ spatial variable for an $N$ electron system, difficulty of calculating it – even approximately – increases with the increasing number of electrons. Thus, attempts were made since the very early days of Quantum Mechanics to develop methods that bypass such calculations in favor of the electronic density $\rho(\mathbf{r})$ because density-based methods require only three space variables irrespective of the number of electrons. This greatly eases the numerical implementation of these methods. The first such method was the Thomas-Fermi method that we discuss next.

## 2. Working in terms of electron density: Thomas-Fermi theory and its extensions [56-59]

We are now moving towards the use of electronic density instead of wavefunction to develop a theory of materials. For this we need to express different quantities of interest in terms of the density. An important question therefore is whether it is possible to formulate a quantum-mechanical theory of materials in terms of electronic density. We will defer answering this question to the next section when we present modern density functional theory. Right now, we focus our attention on the earliest theory of electronic structure in terms of density, known as the Thomas-Fermi theory, and its extensions. Mathematical tools in formulating this theory are the same as those employed in developing modern density-functional theory. This subsection therefore provides the reader a good background for understanding density-functional theory discussed in the rest of the article.

In Thomas-Fermi theory, **kinetic energy per electron** for any system with space-dependent inhomogeneous density $\rho(\mathbf{r})$ is approximated by the formula of Eq. (28a) for HEG. Thus, the total energy of electrons in this theory is a functional of density and is given as

$$E_{TF}[\rho] = \frac{3}{10}(3\pi^2)^{\frac{2}{3}} \int \rho^{\frac{5}{3}}(\mathbf{r}) d\mathbf{r} + \int \rho(\mathbf{r}) v_{ext}(\mathbf{r}) d\mathbf{r} + \frac{1}{2} \iint \frac{\rho(\mathbf{r})\rho(\mathbf{r}')}{|\mathbf{r}-\mathbf{r}'|} d\mathbf{r} d\mathbf{r}' \quad . \quad (52)$$

To get the equation for the density, the energy above is minimized with respect to the density under the condition

$$\int \rho(\mathbf{r}) \, d\mathbf{r} = N \quad , \quad (53)$$

i.e., the density integrates to $N$, the total number of electrons. This leads to the Euler-Lagrange equation (see **supplemental material**)

$$\frac{\delta E[\rho]}{\delta \rho(r)} = \mu \quad , \tag{54}$$

where $\mu$, a constant, is the Lagrange multiplier used to satisfy the condition of Eq. (53). Using the Thomas-Fermi expression for the energy in Eq. (54) gives the equation

$$\frac{1}{2}(3\pi^2\rho(r))^{\frac{2}{3}} + v_{ext}(r) + \int \frac{\rho(r')}{|r-r'|}dr' = \mu \quad , \tag{55a}$$

for the density. This is known as the Thomas-Fermi equation and is to be solved self-consistently to obtain the density.

Let us now understand what physics is implied by the equation above. For this we identify $(3\pi^2\rho(r))^{\frac{1}{3}}$ as the local Fermi wavevector [15,22] $k_F(r)$ at position $r$ and write the Thomas-Fermi equation as

$$\frac{k_F^2(r)}{2} + v_{ext}(r) + \int \frac{\rho(r')}{|r-r'|}dr' = \mu \quad , \tag{55b}$$

From Eq. (55b) it is clear that although the kinetic energy of an electron in the highest occupied orbital varies over different points, its total energy – which is the sum of its kinetic, external and Coulomb energies in presence of other electrons – is a constant throughout the system and is equal to $\mu$. If it was not a constant, the system could lower its energy by reducing the electrons in the regions of higher total energy per electron to the regions of lower energy, making it equal throughout the system when the energy minimum is achieved. Eq. (55b) also leads the physical meaning of $\mu$: it is the energy of the highest energy electrons. Furthermore, negative of $\mu$ is the **removal energy of an electron** from the system, i.e., $\mu$ is the **chemical potential**. This is because if we multiply Eq. (54) by a small change $\Delta\rho(r)$ in the density - corresponding to a change $\Delta N = \int \Delta\rho(r)dr$ in the number of electrons - and integrate over the volume, we get

$$\Delta E = \mu \Delta N \quad , \tag{56a}$$

or

$$\mu = \left(\frac{\partial E}{\partial N}\right)_{v_{ext}} . \tag{56b}$$

Notice that in the equation above, the external potential is kept fixed in taking the derivative. This is a requirement in applying the variational principle and we will come back to this point again in later sections. Equations (55b) and (56b) also indicate that the energy of the highest energy electron is equal to the chemical potential of the system. What is the value of chemical potential itself? That answer is obtained by taking $\Delta N = \pm 1$ and gives

$$\mu = \mp(E_{N\pm1} - E_N) = \begin{cases} -I \text{ for } \Delta N = -1 \\ -A \text{ for } \Delta N = +1 \end{cases} , \tag{57}$$

where $I$ is the ionization potential and $A$ is the electron affinity of the system. This is similar to Koopmans' theorem in Hartree-Fock theory. Our discussion here is depicted schematically for a HEG and an inhomogeneous electron gas in one dimension in **Figure 2**. Since the system taken is very large – like a bulk metal – the orbital energies are almost continuous, and the ionization potential and electron affinity of the system are the same and equal to the **work function $W$**. We remark on the calculation of chemical potential in Eq. (57) that we have done it using finite difference $\Delta N = \pm 1$ while varying the electron number. In part-II of this article, it will be shown rigorously that calculation can also be performed using $\Delta N \to 0$ and that leads the same result.

> **Exercise:** Taking a system of bound electrons to be neutral (due to the background positive charge binding the electrons), evaluate the left-hand side of Eq. (55a) at infinitely large distance from it to obtain the chemical potential of the system in Thomas-Fermi theory. Hence comment on what will the ionization potential of a neutral atom or molecule in Thomas-Fermi theory.

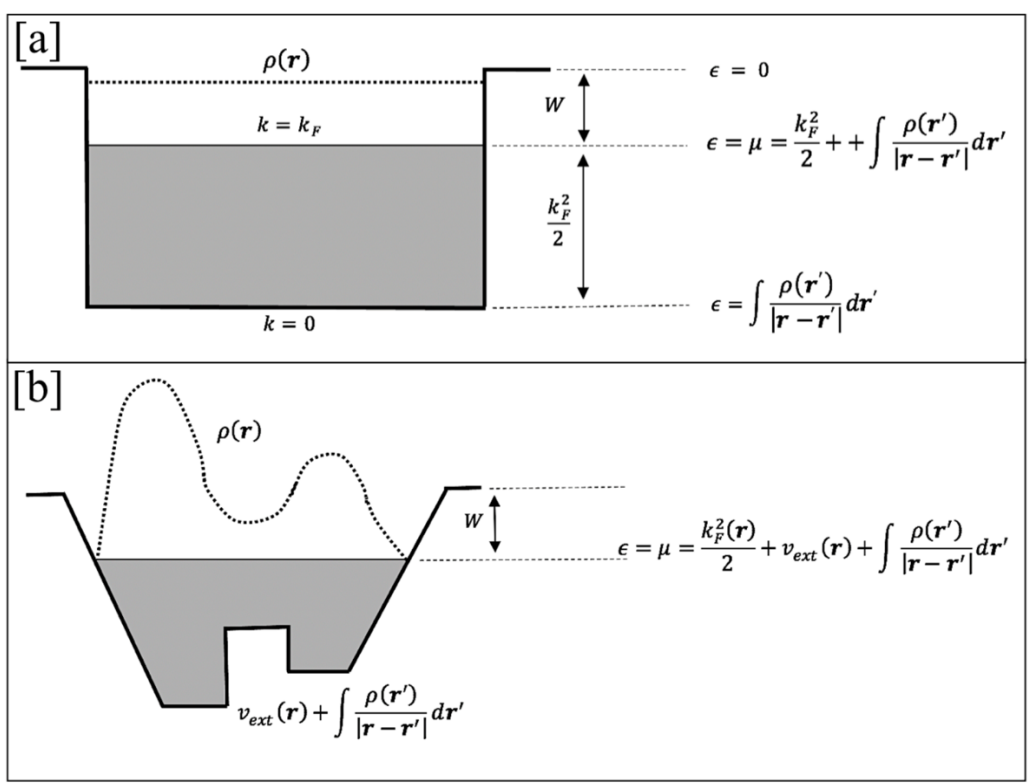

**Figure 2.** (a) Schematic depiction of electrons in a constant potential. The shaded region shows the energy levels filled with the electrons and the dotted line shows the corresponding density of electrons. Note that the kinetic and potential energies of an electron are constant throughout the system; (b) Electrons confined in a varying potential with shaded regions showing the filled levels and dotted line showing the electron density. Notice that the kinetic and potential energies of the highest energy electron now depend on its position in the system. However, its total energy is constant throughout the system.

After doing the exercise in the box above, the reader will realize immediately that Thomas-Fermi theory is missing some essential physics. First, the Hartree potential is calculated for all the electrons and, therefore, an electron has self-interaction in this theory. This can be corrected by subtracting from Thomas-Fermi energy a term which is $N$ times the average self-interaction energy for each electron. This is known as Fermi-Amaldi correction [60] and the total energy with this correction is given as

$$E_{TF}[\rho] - \frac{1}{2N} \iint \frac{\rho(\boldsymbol{r})\rho(\boldsymbol{r}')}{|\boldsymbol{r}-\boldsymbol{r}'|} d\boldsymbol{r} d\boldsymbol{r}' \quad . \quad (58)$$

However, as we learnt in going from Hartree theory to Hartree-Fock theory, Coulomb interaction energy between the electrons is better given by the sum of the Hartree energy and the exchange energy. Considering this, Thomas-Fermi theory is extended more appropriately by including the exchange energy in $E_{TF}[\rho]$. For this purpose, the exchange energy is calculated approximately [61] by using expression (47a) for the exchange energy per electron for HEG and replacing the constant density in it by the space-dependent density $\rho(\boldsymbol{r})$ for an inhomogeneous electron gas. The approximation is similar to that made for the kinetic energy. It is known as the **local-density-approximation (LDA)**. In this approximation, inhomogeneous electron gas is treated as if it is locally homogeneous and any variation in the density at a point does not have significant effect the quantity (kinetic or exchange energy) being approximated. The expression for the exchange energy within the LDA is therefore

$$E_x^{LDA}[\rho] = -\frac{1}{4}\left(\frac{3}{\pi}\right)^{\frac{1}{3}} \int \rho^{\frac{4}{3}}(\boldsymbol{r}) d\boldsymbol{r} \quad . \quad (59)$$

The theory based on inclusion of exchange energy in Thomas-Fermi energy is known as Thomas-Fermi-Dirac (TFD) theory. The total energy in TFD theory is given as

$$E_{TFD}[\rho] = E_{TF}[\rho] + E_x^{LDA}[\rho] \quad . \quad (60)$$

Further improvement in the energy is made by including the correlation energy in the expression above within the LDA by using equation (51b).

A significant shortcoming of Thomas-Fermi theory and its extensions discussed above is the lack of shell structure when density is calculated for atoms. Keep in mind that properties of materials depend on how electrons are distributed among different shells in its constituent atoms. Therefore, shell structure of atoms is important for understanding properties of materials correctly. To get the shell structure, it is necessary to treat kinetic energy of a system in a much better way. This is done by incorporating in the kinetic energy expression terms that depend on the gradient of the density, thus going beyond the LDA and taking into account the inhomogeneity of electron gas. The first such modification to the kinetic energy functional of TF theory is the von-Weizsäcker correction [62]

$$T_W[\rho] = \frac{1}{8} \int \frac{|\nabla \rho(\boldsymbol{r})|^2}{\rho(\boldsymbol{r})} d\boldsymbol{r} \quad . \quad (61)$$

The total energy is now written as the sum of the Thomas-Fermi energy, the exchange energy, the correlation energy and the von-Weizsäcker correction, and the corresponding equation for the density is obtained by applying the variational principle. This theory is known as extended Thomas-Fermi theory. It can be improved further by adding more correction terms to the energy expression.

**Exercise:** Obtain the equation for density in extended Thomas-Fermi theory as more and more energy terms are added to the energy functional.

For an extensive review and application of Thomas-Fermi theory to a variety of systems, the reader is directed to references [58, 59]. As an example of applying extended Thomas-Fermi theory by employing an approximate variational form of the density, we refer the reader to references [63, 64]. Applications of the theory to understand behaviour of bulk materials will be found in references [65, 66].

Having discussed the many-body Schrödinger equation, its solutions and density-based Thomas-Fermi theory, we are now ready to present modern density-functional theory (DFT) [66-74], which is an exact formulation of the many-electron problem in terms of the ground-state density of electrons. The significance of DFT in theory of material design is that it makes solution of the problem much easier than solving the Schrödinger equation. Its implementation, however, requires approximating exchange and correlation energies. Starting from the local-density approximation, these approximations have evolved [75, 76] since the inception of DFT to a high degree of accuracy. This coupled with the ease of employing DFT has made it the most widely used theory of materials. Research in theory of materials therefore delves into understanding fundamental aspects of DFT and developing computational methods of implementing it over a range of materials.

In the next section we begin our discussion of theory of materials by presenting basic formulation of DFT and how it is applied by using the LDA. Numerical techniques to employ it in a variety of systems are also discussed. This will bring us to the end of part-I of the article. Part-II of the article will be devoted to going beyond the LDA with proper understanding of fundamentals of DFT. Section 3 after that is devoted to discussion of some important exact results in DFT.

## 2. Modern DFT

### 2a. Hohenberg-Kohn theorem and connection with approximate density-based theories

In the above, we have discussed Thomas-Fermi theory and its extensions. The theory is based on expressing different energy components in terms of the density of electrons, using mainly the results of HEG. The question arises if the formulation in terms of density is an approximate method only or is it approximation to an exact quantum-mechanical theory in terms of the density. It turns out that one can reformulate many-electron problem in terms of the ground-state of electron density based on **Hohenberg-Kohn (HK) theorems** [77] that we now discuss.

***Theorem I:*** *For $N$ electrons in a multiplicative external potential $v_{ext}(\boldsymbol{r})$, there is one-to-one correspondence between the external potential and the ground-state density $\rho(\boldsymbol{r})$ of electrons.*

Before giving the proof of the theorem, let us explain the question the theorem gives answer to. To determine density of electrons one solves the Schrödinger equation and obtains the density from the resulting wavefunction using Eq. (9). The question that arises is as follows: if the Schrödinger equation is solved with two different multiplicative external potentials in the Hamiltonian, although the wavefunctions will be different, can the ground-state densities still be the same. This question comes up because of the possibility that different antisymmetric functions, when substituted in Eq. (9), can still lead to the same density.

**Exercise:** Show that two different external potentials differing by more than a constant cannot have any common stationary-state wavefunctions.

We give the proof here for non-degenerate ground-states. After that we comment on its generalization [73] for degenerate case.

***Proof:*** *The proof of the theorem is given by contradiction. In giving the proof we restrict ourselves to non-degenerate ground-states. We assume that two different potentials $v_{ext}(\boldsymbol{r})$ and $\tilde{v}_{ext}(\boldsymbol{r})$, whose difference is not a constant, lead to the same ground-state density $\rho(\boldsymbol{r})$ and show that it leads to an absurd result. To do this, assume that the Hamiltonian $H$ with external $v_{ext}(\boldsymbol{r})$ and $\tilde{H}$ with external potential $\tilde{v}_{ext}(\boldsymbol{r})$ have non-degenerate ground-state wavefunctions $\Psi$ and $\tilde{\Psi}$,* *and the corresponding ground-state energies $E$ and $\tilde{E}$, respectively. However, both $\Psi$ and $\tilde{\Psi}$ give the same density $\rho(\mathbf{r})$. By the variational principle*

$$E = \langle \Psi | H | \Psi \rangle < \langle \tilde{\Psi} | H | \tilde{\Psi} \rangle \quad . \tag{62}$$

*Note that the inequality above is strictly less than and not less than or equal to. Now write*

$$H = \tilde{H} + \sum_{i=1}^{N} \left( v_{ext}(\mathbf{r}_i) - \tilde{v}_{ext}(\mathbf{r}_i) \right) \quad , \tag{63}$$

*so that we get*

$$E < \langle \tilde{\Psi} | H | \tilde{\Psi} \rangle = \langle \tilde{\Psi} | \tilde{H} | \tilde{\Psi} \rangle + \int \left( v_{ext}(\mathbf{r}) - \tilde{v}_{ext}(\mathbf{r}) \right) \rho(\mathbf{r}) \, d\mathbf{r} \; ,$$

$$E < \tilde{E} + \int \left( v_{ext}(\mathbf{r}) - \tilde{v}_{ext}(\mathbf{r}) \right) \rho(\mathbf{r}) \, d\mathbf{r} \quad . \tag{64a}$$

*Now use $\tilde{E} = \langle \tilde{\Psi} | \tilde{H} | \tilde{\Psi} \rangle < \langle \Psi | \tilde{H} | \Psi \rangle$ and follow the steps above to get*

$$\tilde{E} < E - \int \left( v_{ext}(\mathbf{r}) - \tilde{v}_{ext}(\mathbf{r}) \right) \rho(\mathbf{r}) \, d\mathbf{r} \tag{64b}$$

*Adding Eqs. (64a) and (64b) gives*

$$E + \tilde{E} < E + \tilde{E} \quad . \tag{65}$$

*This is an absurdity since a quantity cannot be strictly less than itself. Thus, our initial assumption that both the potentials lead to the same ground-state densities is not correct. In other words, each external potential gives a unique ground-state density.*

*The theorem is easily generalized [73] to degenerate case by showing that two degenerate ground-states belonging to two potential that differ by more than a constant cannot be the same. This is done following exactly the same steps as taken in the non-degenerate case. A special situation can, however, also arise in the degenerate case whereby two degenerate wavefunctions give the same density, for example in the non-interacting lithium atom. We will comment on this aspect further in the following.*

With this theorem it is clear that a system with a given number of electrons is fully specified either by the external potential or its ground-state density. As a result, we should be able to obtain any property of an electronic-system from its ground-state electron density. However, the theorem above does not give a prescription for doing this. Contrast this with the case when the external potential is given for a system. In that situation, the path to find any property is straightforward: solve the Schrödinger equation to get the wavefunction and use it to obtain the expectation values of the corresponding operator. So, while the theorem above provides the basis for electronic-structure theory in terms of the density, more work is needed for development of the theory. First step in this direction is the second HK theorem that establishes the variational principle in terms of ground-state density. Before we state and prove this theorem, some mathematical consequences of *theorem I* are presented below.

Because of *theorem I*, the wavefunction of a many-electron system is a functional $\Psi[\rho]$ of its ground-state density $\rho(\boldsymbol{r})$, which means that the ground-state energy is also a functional $E[\rho]$ of the density. We write the energy functional as

$$E_{v_{ext}}[\rho] = \langle \Psi[\rho]|\hat{T} + \hat{V}_{ee}|\Psi[\rho]\rangle + \int \rho(\boldsymbol{r})\, v_{ext}(\boldsymbol{r})d\boldsymbol{r} \quad , \quad (66)$$

where subscript $v_{ext}$ is written explicitly to indicate that that energy is being calculated for the Hamiltonian with external potential $v_{ext}(\boldsymbol{r})$. Notice that in the energy functional above, we have not written $v_{ext}(\boldsymbol{r})$ as dependent on the density. This is to be able to apply the variational principle in terms of the density whereby the Hamiltonian is to be kept fixed when the density is varied. The first term in the expression for energy above is the sum of the kinetic and electron-electron interaction energy and depends on density alone. It is therefore a **universal functional of the density** and its form should be the same for all systems. In the following we will write this functional as $F[\rho]$. Thus,

$$F[\rho] = T[\rho] + E_H[\rho] + E_{xc}^{QM}[\rho] \quad , \quad (67)$$

where the three terms on the right-hand side of the equation above represent, respectively, the kinetic energy, the Hartree energy and the exchange-correlation energy of electrons. We recall that by definition the exchange-correlation energy is

$$E_{xc}^{QM}[\rho] = \langle \Psi[\rho]|\hat{V}_{ee}|\Psi[\rho]\rangle - E_H[\rho] \quad , \quad (68)$$

In Thomas-Fermi theory, $F[\rho]$ is taken to be the sum of the kinetic and Hartree energies (the first and the third terms in Eq. (52)), with the kinetic energy approximated in terms of the density.

For the degenerate case, we want to emphasize that the wavefunction cannot in general be written as a functional of the ground-state density (refer to the example of Li atom above). However, the ground-state energy is still a functional of the density [73] and can be written as the sum of the universal functional of the density and the external energy term.

Next, we present *theorem II*.

**Theorem II:** *The ground-state energy functional attains its minimum value, which is the true ground-state energy, for the correct ground-state density.*

**Proof:** *Consider a system which has external potential $v_{ext}(\boldsymbol{r})$ and ground-state density $\rho(\boldsymbol{r})$. Its ground-state wavefunction is $\Psi[\rho]$. If we now consider some other ground-state density $\rho'(\boldsymbol{r})$ with the corresponding wavefunction $\Psi'[\rho']$. Then by the variational principle for the energy (see Eq. (62)), it follows that*

$$E_{v_{ext}}[\rho] = \langle \Psi[\rho]|\hat{T} + \hat{V}_{ee}|\Psi[\rho]\rangle + \int \rho(\boldsymbol{r})\, v_{ext}(\boldsymbol{r})d\boldsymbol{r}$$

$$< \langle \Psi'[\rho']|\hat{T} + \hat{V}_{ee}|\Psi'[\rho']\rangle + \int \rho'(\boldsymbol{r})\, v_{ext}(\boldsymbol{r})d\boldsymbol{r} = E_{v_{ext}}[\rho'] \quad . \quad (69)$$

The theorem above along with its proof shows that $E_{v_{ext}}[\rho]$ attains its minimum value, which is the true ground-state energy, when the correct ground-state density corresponding to $v_{ext}(\boldsymbol{r})$ is substituted in Eq. (66). For any other density, the value of this expression will be higher. We explicitly mention that in obtaining the inequality above, energy for every density is calculated by keeping $v_{ext}(\boldsymbol{r})$ - and therefore the Hamiltonian - fixed, as should be done in applying the variational principle.

*Theorem II* provides the way to use *theorem I* to develop density-functional theory further. Because the ground-state energy is minimum for the correct ground-state density, we can get the latter by minimizing the energy functional with respect to the density under the constraint of Eq. (53) that the total number of electrons remain unchanged. This leads to the equation

$$\frac{\delta E_{v_{ext}}[\rho]}{\delta \rho(\boldsymbol{r})} = \mu \quad , \qquad (70a)$$

for the ground-state density, where $\mu$ is the Lagrange multiplier used to ensure that the constraint electron number is fixed. Here again, the subscript $v_{ext}$ indicates that when the density is varied in search of the ground-state energy, the external potential is kept fixed, as was noted after presenting the proof of *theorem II*. Eq. (70a) is now written in terms of the functional derivative of the universal functional as

$$\frac{\delta F[\rho]}{\delta \rho(\boldsymbol{r})} + v_{ext}(\boldsymbol{r}) = \mu \quad . \qquad (70b)$$

What we have presented in Eq. (70b) applies equally well to the degenerate ground-states as both the energy and universal functional for these can be written in terms of the density.

We have already come across an approximation to Eq. (70b) in the form of Thomas-Fermi equation: when the universal functional is approximated in Thomas-Fermi theory, Eq. (70b) leads to this equation. The two Hohenberg-Kohn theorems presented above thus provide a rigorous foundation for developing theory of electronic-structure in terms of density and show that Thomas-Fermi theory and its extensions indeed stand on sound mathematical principles. We now comment on the interpretation of the Lagrange multiplier $\mu$ in exact DFT.

Since we have already shown that $\mu$ in Thomas-Fermi theory comes out to be the chemical potential which is equal to the removal energy of an electron from the system, it is no surprise that $\mu$ in the exact theory also has the same meaning. Furthermore, since the theory is exact, the chemical potential of Eq. (70a) or (70b) will be equal to the experimental ionization potential. The proof is exactly the same as given for Thomas-Fermi theory and we urge the reader to go over it once more replacing the approximate energy functionals of Thomas-Fermi theory by the exact ones.

Let us now summarize what has been established. It has been shown that ground-state density of a system can be used as a fundamental variable to describe it. Furthermore, the energy of a system can be written as a functional of this density and is the sum of a universal functional $F[\rho]$ and energy of interaction $\int \rho(\boldsymbol{r}) v_{ext}(\boldsymbol{r}) d\boldsymbol{r}$ with the external potential. Next, the equation for the density has been obtained by applying the variational principle. Solution of this equation gives both the ground-state density and the chemical potential of the system. Thus, if the universal functional is known exactly, the ground-state density, the corresponding energy and the chemical potential of any system can be obtained by performing calculations entirely in terms of density, thereby bypassing the need to obtain the many-body wavefunction. This is precisely what makes density-functional theory charming and operationally effective.

Having established that electronic structure calculation can in principle be performed in terms of density, we now discuss how it is done in practice. One approximate way of applying the theory we have already described in detail is the Thomas-Fermi theory and its extensions. However, as noted there, approximate treatment of kinetic energy leads to many shortcomings in a density-based theory. How does one then treat kinetic energy better while employing density as the basis variable? The answer is provided by Kohn-Sham reformulation of Eqs. (70a) and (70b). We describe that next.

## 2b. The Kohn-Sham Equation

We now discuss the Kohn-Sham formulation [78] of density-functional theory that treats kinetic energy very accurately. This theory is formulated in terms of orbitals of a non-interacting system, called the **Kohn-Sham system,** of the same density as the real system. In this system, the non-interacting electrons move in a multiplicative potential known as the **Kohn-Sham potential**. In proposing such a system, it is assumed that such a potential exists. This assumption is known as the density being noninteracting v-representable. Uniqueness of this potential is guaranteed by the Hohenberg-Kohn theorem. The potential can be obtained [79] from the many-electron wavefunction using the differential virial theorem and has a physical interpretation [80, 81] in terms of classical fields. Orbitals of the KS system are then obtained by solving the corresponding Schrödinger equation, which is known as the **Kohn-Sham (KS) equation**. Practical importance of the formulation, however, arises from the fact that the KS potential can in principle be written exactly in terms of the density. Then the KS equation becomes like the Hartree equation and is solved self-consistently to get the orbitals and the density of a many-electron system. We elaborate on this in the following.

Imagine a system of non-interacting electrons (particles with all properties the same as those of electrons except that they have no charge and are therefore non-interacting) that have the same density as the true system.

> **Comment:** This is an advantage of working with the density: systems with different kinds of inter-particle interaction can have the same ground-state density by adjusting the external potential appropriately. Since the external potentials are different, wavefunctions for each system also differ. A question may be raised: if the external potentials are different, how can the ground-state densities be the same? Wouldn't that be in violation of the Hohenberg-Kohn *theorem I*? The answer is in the negative since *theorem I* states that for a system of particles, different external potentials lead to different ground-state densities. It is implicit in the theorem that inter-particle interaction is fixed, which became self-evident while proving the theorem. In the present case we are changing the inter-particle interaction and adjusting the external potential so that the density remains unchanged for each inter-particle interaction.

Let the potential seen by non-interacting electrons be the Kohn-Sham potential $v_{KS}(\mathbf{r})$. Then the energy of the KS system is

$$E_{KS}[\rho] = T_s[\rho] + \int \rho(\mathbf{r})\, v_{KS}(\mathbf{r}) d\mathbf{r} \quad , \tag{71}$$

where $T_s[\rho]$ is the kinetic energy of non-interacting electrons of the same density. It will be different from the kinetic energy of interacting electrons with density $\rho(\mathbf{r})$ because their wavefunctions and therefore the expectation values of the kinetic energy operator for the two systems are different. The equation for the ground-state density of this system is

$$\frac{\delta T_s[\rho]}{\delta \rho(\mathbf{r})} + v_{KS}(\mathbf{r}) = \mu \quad . \tag{72}$$

Now we look at the corresponding equation for the true system and bring it to the form of Eq. (72). For this the universal functional is written as follows

$$F[\rho] = T[\rho] + E_H[\rho] + E_{xc}^{QM}[\rho]$$
$$= T_s[\rho] + E_H[\rho] + E_{xc}^{DFT}[\rho] \quad , \tag{73}$$

where the difference $T_c[\rho] = T[\rho] - T_s[\rho]$ between the true kinetic energy and the non-interacting kinetic energy for density $\rho(r)$ has been absorbed in the exchange-correlation energy. Thus exchange-correlation energy in density functional theory

$$E_{xc}^{DFT}[\rho] = E_{xc}^{QM}[\rho] + T_c[\rho] \tag{74}$$

is different from that calculated from the many-body wavefunction and includes in it all the many-body effects. With this, the equation for the density is

$$\frac{\delta T_s[\rho]}{\delta \rho(r)} + v_{ext}(r) + \int \frac{\rho(r')}{|r-r'|} dr' + \frac{\delta E_{xc}^{DFT}[\rho]}{\delta \rho(r)} = \mu \quad . \tag{75}$$

A comparison between Eqs. (74) and (75) gives

$$v_{KS}(r) = v_{ext}(r) + \int \frac{\rho(r')}{|r-r'|} dr' + \frac{\delta E_{xc}^{DFT}[\rho]}{\delta \rho(r)} \quad . \tag{76}$$

As noted earlier, the second term on the right-hand side of the equation above is the Hartree potential and is obtained by taking the functional derivative of the Hartree energy with respect to the density. From now on we will denote it as $v_H(r)$. The new term in the equation above is the last term which is the functional derivative of the exchange-correlation energy functional and is known as the exchange-correlation potential and denoted as $v_{xc}(r)$. Thus

$$v_{xc}(r) = \frac{\delta E_{xc}^{DFT}[\rho]}{\delta \rho(r)} \quad . \tag{77}$$

Now Eq. (72) for the density is equivalent to solving the Schrödinger equation for non-interacting electrons and finding the density from the orbitals obtained using Eq. (23). Therefore, Eq. (76) implies that the ground-state density of the interacting system can be found equivalently by solving the Schrödinger-like equation

$$\left(-\frac{1}{2}\nabla^2 + v_{ext}(r) + v_H(r) + v_{xc}(r)\right)\varphi_i(r) = \epsilon_i \varphi_i(r) \quad , \tag{78}$$

to get orbitals $\{\varphi_i(r)\}$, filling the orbitals with lowest possible energies following Pauli's exclusion principle and calculating the density using Eq. (23), i.e.

$$\rho(r) = \sum_{i=1}^{N} |\varphi_i(r)|^2 \quad . \tag{23}$$

Eq. (78) is known as the Kohn-Sham (KS) equation, and the potential $v_{KS}(r)$ and the orbitals $\{\varphi_i(r)\}$ as the KS potential and KS orbitals, respectively. Like the Hartree or the Hartree-Fock equations, KS equation is solved self-consistently since the Hartree and the exchange-correlation potentials depend on the density. Thus, one starts self-consistency iterations with an approximate density, constructs the Hartree and exchange-correlation potentials from it and solve the KS equation to get the orbitals. These orbitals are then used to get the new density using Eq. (23) and this density is used to get new Hartree and exchange-correlation potentials and solve the KS again. These iteration cycles are repeated until the input and output densities match within a chosen tolerance. The density so obtained is the true ground-state density of the system.

Once the orbitals and the density have been obtained, these are used to calculate the energy. The non-interacting kinetic energy of the system is

$$T_s[\rho] = \sum_{i=1}^{N} \left\langle \varphi_i \left| -\frac{1}{2}\nabla^2 \right| \varphi_i \right\rangle \quad . \tag{79}$$

When substituted in the expression for energy, it gives

$$E_{v_{ext}}[\rho] = \sum_{i=1}^{N} \left\langle \varphi_i \left| -\frac{1}{2}\nabla^2 \right| \varphi_i \right\rangle + \int \rho(\mathbf{r})\, v_{ext}(\mathbf{r}) d\mathbf{r} + E_H[\rho] + E_{xc}^{DFT}[\rho] \quad , \tag{80a}$$

The total energy can also be written in terms of the orbital eigen-energies as

$$E_{v_{ext}}[\rho] = \sum_{1}^{N} \epsilon_i - \frac{1}{2}\iint \frac{\rho(\mathbf{r})\rho(\mathbf{r}')}{|\mathbf{r}-\mathbf{r}'|} d\mathbf{r} d\mathbf{r}' - \int \rho(\mathbf{r})\, v_{xc}(\mathbf{r}) d\mathbf{r} + E_{xc}^{DFT}[\rho] \quad . \tag{80b}$$

In passing we note that in terms of the orbital eigen-energies, the energy of the KS system given by Eq. (71) is

$$E_{KS}[\rho] = \sum_{1}^{N} \epsilon_i \quad . \tag{81}$$

As noted earlier, structure of the KS equation and the self-consistent method of solving it are exactly like the Hartree equation (Eq. (26)) and the HF equation (Eqs. (45a) and (45b)). However, there are significant fundamental differences between the KS method and the other two methods. In the KS equation, the exchange-correlation potential is exact while in Hartree theory it is approximated by the self-interaction potential of each orbital. Furthermore, while the KS potential is the same for all orbitals, effective potential (Eq. (27)) in Hartree theory is orbital dependent. Nonetheless, both the exchange-correlation potential and the self-interaction potential are multiplicative. Next, we compare the KS equation and the HF equation. Again, while the KS equation has the exact exchange-correlation potential in it, the HF equation has only the exchange potential. In addition, while the exchange-correlation potential in KS equation is a local potential, i.e., it is multiplicative, the exchange potential in HF theory is non-local. **Local nature of the exchange-correlation potential and its orbital-independence make solving the KS equation easier than the Hartree or the Hartree-Fock equations**. Finally, the most important difference between Hartree or Hartree-Fock theories and Kohn-Sham method is in their philosophy of solving the many-electron problem. The former two are **approximate methods** based on wavefunctional approach and obtain an approximate wavefunction in terms of single-particle orbitals. In contrast, KS formulation is an **exact theory** which in principle gives the true ground-state density of a many-electron system directly, circumventing thereby the requirement of first having to calculate its wavefunction; orbitals in KS theory are just a mathematical construct which lead to this density. Consequently, whereas solutions of Hartree or Hartree-Fock equations can be thought of approximate orbitals of an electron in a many-electron system, no such physical significance can be attributed to orbitals in KS theory. Similarly, orbital eigen-energies in KS theory cannot be interpreted as removal energies - in contrast to orbital energies in Hartree-Fock theory that are approximately the removal energies from the corresponding orbitals - except that for the highest occupied orbital. We elaborate on it in the following.

The Kohn-Sham system, by the way it is constructed, has the same chemical potential $\mu$ as the real system. From Eq. (81), the chemical potential for the KS system is also equal to the orbital eigen-energy $\epsilon_{max}$ of the highest occupied orbital. Since $\epsilon_{max}$ is the removal energy for the KS system, it follows that

$$\epsilon_{max} = -I \quad . \tag{82}$$

Thus, in addition to giving the ground-state density and energy, solution of KS equation also gives the ionization potential of a many-electron system. This is known as the **ionization potential (IP) theorem** of DFT.

In the discussion so far, we have established the exact DFT formalism. We have learnt that in principle the theory leads to the exact ground-state density, energy and the ionization potential. However, implementation of DFT directly in terms of density requires that the kinetic and the exchange-correlation energy functional be approximated. This, as we saw in the context of Thomas-Fermi theory, leads to results that are highly unphysical, primarily because of approximate treatment of kinetic energy in terms of density. However, kinetic energy can be made very accurate if its non-interacting component is treated exactly. This is done in terms of orbitals of an auxiliary system of non-interacting electrons and leads to KS formalism of DFT. Solving the KS equation requires only the exchange-correlation energy to be approximated. As these approximations are made more and more accurate, KS formalism should lead to better and better results. As such, since the inception of DFT, KS formalism has been the mainstay of electronic-structure calculations for material design. In the beginning, KS formalism was applied [82 -84] by employing the LDA for exchange and correlation energies [see Eqs. (47a), (47b), (51b) and (59)] and is the zeroth order approximation which forms the foundation for better approximations developed as the theory evolved. Even today, most of the times, it is the first calculations that is performed on a system before more sophisticated functionals are employed. Next subsection is therefore devoted to discussing the LDA for exchange and correlation and numerical methods employed to solve the KS equation using this approximation.

## 2c. Solving the Kohn-Sham equation by employing the LDA

In this subsection, we discuss how the KS equation is solved by employing the LDA for the exchange-correlation energy and therefore also the exchange-correlation potential. The functional form for the exchange energy in the LDA has been given in Eq. (59). In general, exchange-correlation energy in the LDA is written as

$$E_{xc}^{LDA}[\rho(\mathbf{r})] = \int \rho(\mathbf{r}) \epsilon_{xc}^{HEG}(\rho(\mathbf{r})) \, d\mathbf{r} \tag{83}$$

where $\epsilon_{xc}^{HEG} (= \epsilon_{x}^{HEG} + \epsilon_{c}^{HEG})$ is the exchange-correlation energy per electron for charge density $\rho$ of homogeneous electron gas. Exact expression for $\epsilon_{x}^{HEG}$ is given in Eq. (47a) and a formula based on interpolation for $\epsilon_{c}^{HEG}$ is given in Eq. (51b). The potentials are calculated by taking the functional derivative (see **supplemental material**) of the corresponding functionals and are given as

$$v_{x}^{LDA}(\mathbf{r}) = -\left(\frac{3\rho(\mathbf{r})}{\pi}\right)^{1/3} \quad , \tag{84a}$$

and

$$v_c^{LDA}(\mathbf{r}) = -\frac{0.147}{(r_s(\mathbf{r}) + 7.85)^2}(4r_s(\mathbf{r}) + 23.55) \quad , \tag{84b}$$

where $r_s(\mathbf{r})$ is related to the density $\rho(\mathbf{r})$ through Eq. (29). While we have used the Wigner interpolation formula for the correlation energy per electron in a HEG, other more accurate expressions for it exist. These are parametrizations given by Hedin and Lundqvist [85], von Barth and Hedin [86], Vosko, Wilk and Nusair [87], and Perdew and Zunger [88] parametrization of Monte-Carlo calculations of Ceperley and Alder [54]. For some recent work a different way of parametrizing the correlation energy for HEG, we refer the reader to ref. [89]. Furthermore, a critical investigation of different parametrization for correlation energy of HEG has been given in ref. [90].

A related approximation to the LDA is the local spin density approximation (LSDA) that calculates the exchange-correlation energy in terms of the densities $\rho_\uparrow(\mathbf{r})$ and $\rho_\downarrow(\mathbf{r})$ of electrons for spins $+\frac{1}{2}$ and $-\frac{1}{2}$, respectively. The energy expression in this approximation is given as

$$E_{xc}^{LSDA}[\rho_\uparrow(\mathbf{r}), \rho_\downarrow(\mathbf{r})] = \int \rho(\mathbf{r}) \epsilon_{xc}^{HEG}(\rho_\uparrow(\mathbf{r}), \rho_\downarrow(\mathbf{r})) \, d\mathbf{r} \tag{85}$$

> **Exercise:** Show that the exchange energy in terms of $\rho_\uparrow(\mathbf{r})$ and $\rho_\downarrow(\mathbf{r})$ can be written as
>
> $$E_x[\rho_\uparrow(\mathbf{r}), \rho_\downarrow(\mathbf{r})] = \frac{1}{2} E_x[2\rho_\uparrow(\mathbf{r})] + \frac{1}{2} E_x[2\rho_\downarrow(\mathbf{r})] \quad . \tag{86}$$
>
> Here the exchange energy on the right-hand side is calculated using the expression for system having equal number of electrons for both spins. Furthermore, show that the non-interacting kinetic energy also follows the same relation.

We note that the correlation energy cannot be split neatly into components dependent on $\rho_\uparrow(\mathbf{r})$ and $\rho_\downarrow(\mathbf{r})$ separately. Nonetheless, in the LSDA the correlation energy too is written in a manner similar to the exchange energy as follows. Since the correlation energy for HEG is calculated for systems where all electrons have the same spin (ferromagnetic and denoted with superscript $F$) or half of the electrons have $+\frac{1}{2}$ and the other half have spin $-\frac{1}{2}$ (paramagnetic and denoted with superscript $P$), the exchange-correlation energy per particle appearing in Eq. (86) is interpolated [77] in terms of density $\rho = \rho_\uparrow + \rho_\downarrow$ and spin polarization parameter $\zeta = \frac{\rho_\uparrow - \rho_\downarrow}{\rho_\uparrow + \rho_\downarrow}$ as

$$\epsilon_{xc}^{HEG}(\rho, \zeta) = \epsilon_{xc}^P(\rho, 0) + [\epsilon_{xc}^F(\rho, 1) - \epsilon_{xc}^P(\rho, 0)] f(\zeta) \quad , \tag{87a}$$

where

$$f(\zeta) = \frac{1}{2}\left[\frac{(1+\zeta)^{4/3} + (1-\zeta)^{4/3} - 2}{2^{1/3} - 1}\right] \quad . \tag{87b}$$

The corresponding potential for electrons of spin $\sigma$ (↑ or ↓) in the LSDA is also written following the exact form for the exchange potential and is given as [86,88]

$$v_{xc,\sigma}^{LSDA}(\boldsymbol{r}) = v_{xc}^{P}(\rho(\boldsymbol{r}),0) + [v_{xc}^{F}(\rho(\boldsymbol{r}),1) - v_{xc}^{P}(\rho(\boldsymbol{r}),0)]f(\zeta)$$

$$+ [\epsilon_{xc}^{F}(\rho(\boldsymbol{r}),1) - \epsilon_{xc}^{P}(\rho(\boldsymbol{r}),0)][\mathrm{sgn}(\sigma) - \zeta]\frac{df(\zeta)}{d\zeta} \quad , \qquad (88)$$

where $\mathrm{sgn}(\uparrow) = 1$ and $\mathrm{sgn}(\downarrow) = -1$. The LSDA is more accurate than the LDA and is necessary when we consider a magnetic system.

**Exercise:** Show that relations given by Eq. (87a) and Eq. (88) are exact for, respectively, the exchange energy per electron and exchange potential in a HEG.

Using potentials given by Eqs. (84), KS equation can be easily solved numerically for spherically symmetric systems such as atoms [82], jellium spheres [63,91,92] and Hookium [7]. A good guide to writing the numerical code to solve the KS equation self-consistently for spherical systems is in the book by Hermann and Skillman [93]. Solutions of the KS equation give the orbitals and their eigenvalues. From these the density of the system is obtained from Eq. (23) and the total energy from Eq. (80a) or (80b). Furthermore, eigenvalue for the highest energy orbital should give the ionization energy of the system. Results obtained for noble gas atoms He, Ne and Ar by using the LDA are shown in **Table 4.** (within the LDA, the extra electron in the hydrogen negative ion – and other negative ions – does not bind [94, 95]).

**Table 4.** Total energy and eigenenergy of the highest occupied (HO) orbital of three noble gas atoms as calculated within the LDA in comparison with the experimental total energies and ionization potentials. All numbers are given in atomic units.

| Atom | Total energy | | Eigenvalue (HO) and Ionization potential | |
|---|---|---|---|---|
| | $E_{LDA}$ | $E_{expt}$[30, 31] | $-\epsilon_{max}$ | $I_{expt}$ [30, 31] |
| He | -2.8335 | -2.9034 | 0.5696 | 0.9035 |
| Ne | -128.2210 | -129.0600 | 0.4961 | 0.7925 |
| Ar | -525.9280 | -529.2490 | 0.3822 | 0.5790 |

Numbers for total energies in **Table 4**. show that the LDA is a reasonably good approximation and leads to total energies close to the experimental numbers. Approximate treatment of exchange and correlation effects, however, makes them less negative than the corresponding Hartree-Fock energies given in **Table 2**, although correlation energy is included in the total energy calculation. Nonetheless, it is clear that LDA based calculations can be used to obtain the total energies of a system. Furthermore, accuracy of the energies indicates that better approximations for exchange and correlation can be developed with the LDA as the starting point.

In contrast to the total energies, the highest occupied orbital eigen-energies are much smaller in magnitude compared to the experimental ionization energies. This means electrons in these orbitals are not as tightly bound as they should be. The reason for this is that the eigenvalue for the uppermost orbital depends crucially on the potential in the outer regions of a system; the LDA potential in these regions is not as deep as the true potential (exact behaviour of KS potential in asymptotic regions of a system is discussed in part-II of this article). This is easily understood from the exchange potential which varies as $\rho^{1/3}$ and therefore decays exponentially in the outer regions where the density itself varies exponentially with the distance from the system. This leads to weak binding of electrons in the highest occupied orbital.

Next, we present the results for some solids obtained from the solutions of the KS equation by applying the LDA. As mentioned earlier in the article, plane-wave basis has been most popular [96-98] to solve Kohn-Sham equation for solids. Results of KS-LDA calculations are displayed in **Table 5**. It is seen that the lattice constants for solids of C, Si and Ge are given quite accurately by the LDA whereas those for Ne, Ar, Kr and Xe do not have the same accuracy. This is a reflection of the LDA not describing outer low-density regions of these atoms well. The band-gaps for all systems, on the other hand, are underestimated by a significantly large amount by the LDA. We will learn in part-II of the article that underestimation of the gap is not limited to the LDA but even highly accurate approximations give a gap smaller than the true gap. An understanding of why this happens will also be discussed there. Finally, we point out that the difference in the values of energy gaps given in references [101] and [102] arise due to inclusion of relativistic effects in work of ref. [102].

In this subsection we have given some representative results for finite and infinitely large systems by solving the KS equation employing the LDA. These results show that reasonably accurate estimates can be made for many properties of these systems. More importantly, they give the hope that with better approximations for the exchange-correlation energy functionals, more accurate results can be obtained.

**Table 5.** Lattice constants, bulk modulus and energy band gap for some non-metallic systems calculated in LDA. These are compared with the corresponding experimental numbers.

| Solid | $a_{LDA}$ [Å] | $a_{Expt}$ [Å] | $B_{LDA}$ [GPa] | $B_{Expt}$ [GPa] | $E_{Gap}^{LDA}$ [eV] | $E_{Gap}^{Expt}$ [eV] |
|---|---|---|---|---|---|---|
| C | 3.55 [99] | 3.57 [22] | 458 [99] | 442 [34] | 4.10 [100] | 5.50 [22] |
| Si | 5.41 [99] | 5.43 [22] | 95 [99] | 99 [34] | 0.47 [100] | 1.17 [22] |
| Ge | 5.63 [99] | 5.65 [22] | 70 [99] | 76 [34] | 0.00 [100] | 0.74 [22] |
| Ne | 3.86 [101] | 4.46 [37] | -- | 1.1 [38] | 11.32 [101] 11.40 [102] | 21.48 [41] |
| Ar | 4.95 [101] | 5.30 [42] | -- | 2.9 [38] | 8.16 [101] 8.10 [102] | 14.15 [41] |
| Kr | 5.36 [101] | 5.65 [43] | -- | 3.3 [38] | 6.47 [101] 6.76 [102] | 11.59 [41] |
| Xe | 5.90 [101] | 6.13 [44] | -- | 3.7 [38] | 5.26 [101] 5.56 [102] | 9.29 [41] |

## 3. Concluding remarks:

In this article, we have started with an introduction to the Schrödinger equation and presented methods of its solution for simple cases when particles (electrons) are not interacting. It was then discussed in detail how the introduction of interaction between electrons makes solving the equation impossible. Thus, approximate and yet accurate methods of solution have to be developed. In this connection Hartree and Hartree-Fock methods were presented in detail, partly for the reason that the structure of equations to be solved in these methods is similar to the KS equation of DFT. Initial attempts to find an alternative to solving the Schrödinger equation for the wavefunction led to approximate density-based theories starting with Thomas-Fermi theory. This theory was presented in detail as it forms the basis of exact density functional theory and many concepts introduced here appear again when density-functional theory is discussed. Following Thomas-Fermi theory and its extensions, we moved on to describe modern DFT in its exact form both in terms of density and its Kohn-Sham version in terms of orbitals. Finally, we described how Kohn-Sham equation can be solved by making the local-density approximation and its spin dependent version called the local spin density approximation. Result obtained by this method were presented.

To make further progress in applying DFT and get more accurate results, it is important that fundamental aspects of the theory be understood well and are used to make better estimates of properties of materials. These aspects of DFT will be presented in the second part of this article.

**Disclosure statement:** The authors thank the editors for invitation to write the article presented here. MKH is thankful to Rabeet Singh and Ashish Kumar for stimulating discussions and providing some of the results presented here. Comments on the manuscript by Vishal Agrawal are also appreciated. Work at Ames Laboratory was supported by U.S. Department of Energy (DOE) Office of Science, Basic Energy Sciences, Materials Science & Engineering Division. Ames Laboratory is operated by ISU for the U.S. DOE under contract DE-AC02-07CH11358.

**Conflict of interest**: Author's declare no conflict of interest.

**Author contributions**: **PS**: Data curation, Investigation, formal analysis, Writing – original draft, Writing – review & editing. **MKH**: Conceptualization, Supervision, Data curation, Investigation, formal analysis, Writing – original draft, Writing – review & editing.

**References:**

1. R. Eisberg and R. Resnick, *Quantum Physics of Atoms, Molecules, Solids, Nuclei and Particles* (Wiley, 2006) Second Edition.
2. L.I. Schiff, *Quantum Mechanics* (McGraw Hill, 1968) Third Edition
3. A. Haug, Theoretical Solid State Physics Volume 1 (Pergamon, 1972).
4. B.T. Sutcliffe and R.G. Woolley, On the quantum theory of molecules, J. Chem. Phys. **137**, 22A544 (2012).
5. A.L. Fetter and J.D. Walecka, *Quantum theory of many-particle systems* (Dover, New York, 2002).
6. S. Kais, D.R. Herschbach, N.C. Handy, C.W. Murray and G.J. Laming, Density functionals and dimensional renormalization for an exactly solvable model, J. Chem. Phys. **99** 417 (1993)
7. P. M. Laufer and J. B. Krieger, Test of density-functional approximations in an exactly soluble model, Phys. Rev. A **33**, 1480 (1986)
8. K. Frankowski and C.L. Pekeris, Logarithmic Terms in the Wave Functions of the Ground State of Two-Electron Atoms, Phys. Rev. **146**, 46 (1966); Errata Phys. Rev. **150**, 366(E) (1966).
9. C.J. Umrigar and X. Gonze, Accurate exchange-correlation potentials and total-energy components for the helium isoelectronic series, Phys. Rev. A **50**, 3827 (1994)
10. T. Koga, Y. Kasai and A.J. Thakkar, Int. J. Quantum Chem. **46**, 689 (1993)
11. H.A. Bethe and E.E. Salpeter, *Quantum Mechanics of one- and two-electron atoms* (Springer, 2014)
12. C. Le Sech, Correlated wavefunctions for two-electron systems using new screened hydrogen-like orbitals, J. Phys. B: At. Mol. Opt. Phys. **30**, L47 (1997).
13. T.D.H. Baber and H.R. Hasse, A comparison of wave functions for the normal helium atom, Math Proc. Camb. Philos. Soc. **33**, 253 (1937).
14. R. Chauhan and M.K. Harbola, Improved Le Sech wavefunctions for two-electron atomic system, Chem. Phys. Lett. **639**, 248 (2015).
15. N. Ashcroft and N.D. Mermin, *Solid State Physics* (Saunders College, 1976).
16. E.A. Hylleraas, Über den Grundzustand des Heliumatoms, Z. Phys. **48**, 469 (1928); ibid Neue Berechnung der Energie des Heliums im Grundzustande, sowie des tiefsten Terms

von Ortho-Helium, Z. Phys. **54**, 347 (1929); ibid Die Ionisierungsspannungen von Atomkonfigurationen mit zwei Elektronen, Naturwissenschaften **17**, 982 (1929).

17. S. Chandrasekhar, Some Remarks on the Negative Hydrogen Ion and its Absorption Coefficient, Astrophys. J. **100**, 176 (1944).
18. A. Kumar, R. Singh and M.K. Harbola, Accurate effective potential for density amplitude and the corresponding Kohn–Sham exchange–correlation potential calculated from approximate wavefunctions, *J. Phys. B: At. Mol. Opt. Phys.* **53**,165002 (2020).
19. D.R. Hartree, Math. Proc. Cambridge Phil. Soc. **24**, 89 (1928); *ibid* 111 (1928); *ibid* 246 (1928)
20. D.R. Hartree, *The calculation of atomic structures* (John Wiley and Sons, New York, 1957).
21. I.M. Gelfand and S.V. Fomin, *Calculus of Variations* (Dover, 2000)
22. C. Kittel, *Introduction to Solid-state physics* (Wiley, 1996) Seventh Edition.
23. J.C. Slater, *Quantum theory of atomic structure*, Vol. I and II (McGraw Hill, New York, 1960).
24. A. Szabo and N.S. Ostlund, *Modern Quantum Chemistry* (Dover, 1996).
25. M.K. Harbola and V. Sahni, Theories of electronic structure in the Pauli-correlated approximation, J. Chem. Ed. **70**, 920 (1993).
26. T. Koopmans, Über die Zuordnung von Wellenfunktionen und Eigenwerten zu den Einzelnen Elektronen Eines Atom, Physica **1**, 104 (1934).
27. C. F. Fischer, *Hartree-Fock method for atoms. A numerical approach*. United States: N. p., 1977. Web.
28. C. F. Fischer, General Hartree-Fock program, Computer Phys. Comm. **43**, 355 (1987).
29. E. Clementi and C. Roetti, Roothaan-Hartree-Fock atomic wavefunctions: Basis functions and their coefficients for ground and certain excited states of neutral and ionized atoms, Z≤5, At. Data Nucl. Data Tables **14**, 177 (1974).
30. K. R. Lykke, K. K. Murray and W. C. Lineberger, Threshold photodetachment of H-, Phys. Rev. A **43**, 6104 (1991).
31. A. Kramida, Yu. Ralchenko, J. Reader and NIST ASD Team, NIST Atomic Spectra Database (ver. 5.8, 2020); Available online: physics.nist.gov/asd (2021), National Institute of Standards and Technology, Gaithersberg (USA)
32. G.T. Surratt, R.N. Euwema, and D.L. Wilhite, Hartree-Fock Lattice Constant and Bulk Modulus of Diamon, Phys. Rev. B **8**, 4019 (1973).
33. M. Causà, R. Dovesi, and C. Roetti, Pseudopotential Hartree-Fock study of seventeen III-V and IV-IV semiconductors, Phys. Rev. B **43**, 11937 (1991).
34. Numerical Data and Functional Relationships in Science and Technology, edited by O. Madelung, Landolt-Bornstein, New Series, Vol. 17a (Springer, Berlin, 1982)
35. N.E. Brener, Correlated Hartree-Fock energy bands for diamond, Phys. Rev. B **11**, 929 (1975).
36. A. Svane, Hartree-Fock band-structure calculations with the linear muffin-tin-orbital method: Application to C, Si, Ge, and α-Sn, Phys. Rev. B **35**, 5496 (1987).
37. D.N. Batchelder, D.L. Losee, and R.O. Simmons, Measurements of Lattice Constant, Thermal Expansion, and Isothermal Compressibility of Neon Single Crystals, Phys. Rev. **162**, 767 (1967).
38. M. S. Anderson and C. A. Swenson, Experimental equations of state for the rare gas solid, J. Phys. Chem. Solids **36**, 145 (1975).
39. L. Dagens and F. Perrot, Hartree-Fock Band Structure and Optical Gap in Solid Neon and Argon, Phys. Rev. B **5**, 641 (1972).
40. A.B. Kunz and D.J. Mickish, Study of the Electronic Structure and the Optical Properties of the Solid Rare Gases, Phys. Rev. B **8**, 779 (1973)
41. S. Bernstorff, and V. Saile, Experimental determination of band gaps in rare gas solids, Opt. Commun. **58**, 181-186 (1986).
42. O.G. Peterson, D.N. Batchelder, and R.O. Simmons, Measurements of X-Ray Lattice Constant, Thermal Expansivity, and Isothermal Compressibility of Argon Crystals, Phys. Rev. **150**, 703 (1966).

43. D.L. Losee and R.O. Simons, Thermal-Expansion Measurements and Thermodynamics of Solid Krypton, Phys. Rev. **172**, 944 (1968).
44. D.R. Sears and H.P. Klug, Density and Expansivity of Solid Xenon, J. Chem. Phys. **37**, 3002 (1962).
45. D. Bohm and D. Pines, A Collective Description of Electron Interactions: III. Coulomb Interactions in a Degenerate Electron Gas, Phys. Rev. 92, 609 (1953).
46. J. Hubbard, The description of collective motions in terms of many-body perturbation theory. II. The correlation energy of a free-electron gas, *Proc. Royal Soc. of London. Series A. Mathematical and Physica, Sciences*. **243**, 336 (1958).
47. H. Ehrenreich and M.H. Cohen, Self-Consistent Field Approach to the Many-Electron Problem, Phys. Rev. 158, 786 (1959).
48. R. Pariser and R. Parr, A Semi-Empirical Theory of the Electronic Spectra and Electronic Structure of Complex Unsaturated Molecules. I, J. Chem. Phys. **21**, 466, (1953).
49. R. Pariser and R. Parr, A Semi-Empirical Theory of the Electronic Spectra and Electronic Structure of Complex Unsaturated Molecules. II, J. Chem. Phys. **21**, 767 (1953).
50. J. A. Pople, Electron interaction in unsaturated hydrocarbons, Trans. Faraday Soc. **49**, 1375 (1953).
51. J. Hubbard, *Proc. Royal Soc. of London. Series A. Mathematical and Physical Sciences*. **276**, 238 (1963)
52. E. P. Wigner, On the Interaction of Electrons in Metals, Phys. Rev. **46**, 1002 (1934).
53. T. Smoleński et al., Signatures of Wigner crystal of electrons in a monolayer semiconductor, Nature **595**, 53 (2021).
54. D.M. Ceperly and B.J. Alder, Ground State of the Electron Gas by a Stochastic Method, Phys. Rev. Lett. **45**, 566 (1980).
55. E. P. Wigner, Effects of the electron interaction on the energy levels of electrons in metals, Trans. Faraday Soc. **34**, 678 (1938).
56. L.H. *Thomas, The calculation of atomic fields,* Proc. Camb. Phil. Soc*. **23**, 542 (1927).*
57. *E. Fermi, Rend. Un Metodo Statistico per la Determinazione di alcune Prioprietà dell'Atom, Accad. Naz. Lincei. **6**, 602(1927).*
58. N.H March *in Theory of inhomogeneous electron gas*, edited by S. Lundquist and N.H. March (Plenum, New York, 1983)
59. L. Spruch, Pedagogic notes on Thomas-Fermi theory (and on some improvements): atoms, stars, and the stability of bulk matter, Rev. Mod. Phys. **63**, 151 (1991).
60. E. Fermi and E. Amaldi, Accad. Ital. Rome **6**, 119 (1934).
61. P.A.M. Dirac, Note on Exchange Phenomena in the Thomas Atom Show affiliation, Math. Proc. Cambridge Philos. Soc. **26**, 376 (1930).
62. C.F. von Weizsacker, Zur Theorie der Kernmassen, Z. Physik **96**, 431 (1935).
63. M. Brack, The physics of simple metal clusters: self-consistent jellium model and semiclassical approaches, Rev. Mod. Phys. **65**, 677 (1993).
64. J.C. Slater and H. M. Krutter, The Thomas-Fermi Method for Metals, Phys. Rev. **47**, 559 (1935).
65. R. P. Feynman, N. Metropolis, and E. Teller, Equations of State of Elements Based on the Generalized Fermi-Thomas Theory, Phys. Rev. **75**, 1561 (1949).
66. M.K. Harbola and A. Banerjee, An ab initio force field for the cofactors of bacterial photosynthesis, J. Th. Comp. Chem. **02**, 301 (2003).
67. R.O. Jones, Density functional theory: Its origins, rise to prominence, and future, Rev. Mod. Phys. **87**, 897 (2015).
68. K. Burke, Perspective on density functional theory, J. Chem. Phys. **136**, 150901 (2012).
69. N. Argaman and G. Makov, Density functional theory: An introduction, Am. J. Phys. **68**, 69 (2000).
70. R.O. Jones and O. Gunnarsson, The density functional formalism, its applications and prospects, Rev. Mod. Phys. **61**, 689 (1989).
71. E. Engel and R.M. Dreizler, *Density functional Theory, An Advanced Course* (Springer, Berlin/Heidelberg, 2011)

72. N.H. March, *Electron Density Theory of Atoms and Molecules* (Academic, London/New York, 1992).
73. R.M. Dreizler and E.K.U. Gross, *Density Functional Theory* (Springer, Berlin/Heidelberg, 1990).
74. R.G. Parr and W. Yang, *Density-functional theory of atoms and molecules* (Oxford, New York, 1989).
75. R. Peverati and D.G. Truhlar, Quest for a universal density functional: the accuracy of density functionals across a broad spectrum of databases in chemistry and physic, Phil. Trans. Royal Soc. A **372**, 20120476 (2014).
76. M.G. Medvedev, I.S. Bushmarinov**,** J. Sun, J.P. Perdew, K.A. Lyssenko, Density functional theory is straying from the path toward the exact functional, Science **355**, 49 (2017).
77. Hohenberg, W. Kohn, Inhomogeneous Electron Gas, Phys. Rev. **136**, B864 (1964).
78. W. Kohn and L.J. Sham, Self-Consistent Equations Including Exchange and Correlation Effects, Phys. Rev. **140**, A1133 (1965).
79. A. Holas and N.H. March, Exact exchange-correlation potential and approximate exchange potential in terms of density matrices, Phys. Rev. A **51**, 2040 (1995).
80. M.K. Harbola and V. Sahni, Quantum-Mechanical Interpretation of the Exchange-Correlation Potential of Kohn-Sham Density-Functional Theory, Phys. Rev. Lett. 62, 489 (1989).
81. V. Sahni, *Quantal Density Functional Theory*, Second Edition (Springer, Berlin, 2016)
82. B. Y. Tong and L. J. Sham, Application of a Self-Consistent Scheme Including Exchange and Correlation Effects to Atoms, Phys. Rev. **144**, 1 (1966).
83. Computational methods in band theory, Edited by P. M. Marcus, J. F. Janak, and A. R. Williams (Plenum, New York, 1971).
84. B.Y. Tong, Kohn-Sham Self-Consistent Calculation of the Structure of Metallic Sodium, Phys. Rev. B **6**, 1189 (1972).
85. L. Hedin and B.I. Lundquist, Explicit local exchange-correlation potentials, J. Phys. C **4**, 2064 (1971).
86. U. von Barth and L. Hedin, A local exchange-correlation potential for the spin polarized case. I, J. Phys. C **5**, 1629 (1972).
87. S.H. Vosko, L. Wilk and M. Nusair, Accurate spin-dependent electron liquid correlation energies for local spin density calculations: a critical analysis, Can. J. Phys. **58**, 1200 (1980).
88. J.P. Perdew and A. Zunger, Self-interaction correction to density-functional approximations for many-electron systems, Phys. Rev. B **23**, 5048 (1981).
89. J. Sun, J.P. Perdew and M. Seidl, Correlation energy of the uniform electron gas from an interpolation between high- and low-density limits, Phys. Rev. B **81**, 85123 (2010); Erratum Phys. Rev. B **98**, 079903 (2018).
90. P. Bhattarai, A. Patra, C. Shahi, and J. P. Perdew, How accurate are the parametrized correlation energies of the uniform electron gas? Phys. Rev. B **97**, 195128 (2018).
91. F. Sottile and P. Ballone, Fixed-node diffusion Monte Carlo computations for closed-shell jellium sphere, Phys. Rev. B **64**, 045105 (2001).
92. L. M. Almeida, J.P. Perdew and C. Fiolhais, Surface and curvature energies from jellium spheres: Density functional hierarchy and quantum Monte Carlo, Phys. Rev. B **66**, 075115 (2002).
93. F. Herman and S. Skillman, *Atomic Structure Calculations* (Prentice Hall, 1963).
94. H.B. Shore, J.H. Rose and E. Zaremba, Failure of the local exchange approximation in the evaluation of the H− ground state, Phys. Rev. B **15**, 2858 (1977).
95. K. Schwarz, First ionisation potentials of atoms obtained with local-density schemes, J. Phys. B **11**, 1339 (1978); *ibid* Instability of stable negative ions in the Xα method or other local density functional schemes, Chem. Phys. Letters **57**, 605 (1978).
96. G. Kresse, J. Furthmüller, Efficient iterative schemes for ab initio total-energy calculations using a plane-wave basis set, Phys. Rev. B **54**, 11169 (1996).
97. M.D. Segall, P.J.D. Lindan, M.J. Probert, C.J. Pickard, P.J. Hasnip, S.J. Clark, M.C. Payne, First-principles simulation: ideas, illustrations and the CASTEP code, J. Phys. Condens. Matter **14**, 2717-2744 (2002).

98. X. Gonze, J.-M. Beuken, R. Caracas, F. Detraux, M. Fuchs, G.-M. Rignanese, L. Sindic, M. Verstraete, G. Zerah, F. Jollet, M. Torrent, A. Roy, M. Mikami, Ph. Ghosez, J.-Y. Raty, D.C. Allan, First-principles computation of material properties: the ABINIT software project, Comput. Mater. Sci. **25**, 478-492 (2002).
99. Guo-Xu Zhang et al, Performance of various density-functional approximations for cohesive properties of 64 bulk solids, New J. Phys. **20**, 063020 (2018).
100. F. Tran, and P. Blaha, Importance of the Kinetic Energy Density for Band Gap Calculations in Solids with Density Functional Theory, J. Phys. Chem. A **121**, 3318–3325 (2017).
101. R. J. Magyar, A. Fleszar, and E. K. U. Gross, Exact-exchange density-functional calculations for noble-gas solids, Phys. Rev. B **69**, 045111 (2004).
102. N. C. Bacalis, D. A. Papaconstantopoulos and E. Pickett, Systematic calculations of the band structures of the rare-gas crystals neon, argon, krypton, and xenon, Phys. Rev. B **38**, 6218 (1988).

**Supplemental Material**

# Density Functional Theory of Material Design:
# Fundamentals and Applications - I

Prashant Singh[1] and Manoj K. Harbola[2,*]

[1]Ames Laboratory, U.S. Department of Energy, Iowa State University, Ames, Iowa 50011 USA

[2]Department of Physics, Indian Institute of Technology, Kanpur
Kanpur – 208016, India

**Functionals and functional derivatives**

A functional $F[f]$ of a function $f(r)$ of a function maps the function to a number, just as a function assigns a number to another number. Thus, a functional can be

$$F[f] = f(0) \quad , \qquad (S1)$$

i.e, it returns the value of the function at the origin. For example, this functional will give $F[\sin x] = 0, F[\cos x] = 1$ and $F[ax + b] = b$. As the reader must have noted by now, a functional is denoted by putting the function inside square brackets. Generally, functionals that we use in the work presented in this paper are written in the form of an integral as

$$F[f] = \int w(\boldsymbol{r}) g(f(\boldsymbol{r}), \nabla f(\boldsymbol{r}), \nabla^2 f(\boldsymbol{r}) \cdots) d\boldsymbol{r} \quad , \qquad (S2)$$

where $g$ is a function of $f$ and its derivatives, and $w(\boldsymbol{r})$ is a given function. In the example of Eq. ($A1$), $w(\boldsymbol{r}) = \delta(\boldsymbol{r})$ and $g(f) = f$. In general, a functional could have more than one function as its argument. In the context of the variational principle for the energy used in quantum mechanics, all the energy components, and therefore the total energy, are functionals of the wavefunction and its complex conjugate. For a single particle having the wavefunction $\varphi(\boldsymbol{r})$, the kinetic energy is given as the functional

$$T[\varphi^*,\varphi] = -\frac{1}{2}\int \varphi^*(r)\nabla^2\varphi(r)\,dr = \frac{1}{2}\int |\nabla\varphi(r)|^2\,dr \quad , \quad (S3)$$

where the second equality follows using the divergence theorem and the boundary condition that the wavefunction vanishes at the boundary as $|r| \to \infty$. Thus, kinetic energy can be written in the functional form by using either the second derivative or first derivative of the wavefunction. If the particle is in the potential $v_{ext}(r)$, its total energy functional is

$$E[\varphi] = \frac{1}{2}\int |\nabla\varphi(r)|^2\,dr + \int v_{ext}(r)\,|\varphi(r)|^2\,dr \quad . \quad (4)$$

For more than one particles, the wavefunction will be a function of more than 3 coordinates but the mathematic remains the same. A functional can also map a function to another function. An example of this functional relationship is given below

$$f_1(r,[f]) = \int w(r,r')g(f(r'),\nabla' f(r'),\nabla'^2 f(r')\cdots)\,dr' \quad , \quad (S5)$$

To make a mental picture of a functional, it helps to think of it as a function of multiple variables – these variables being the values $f_i$ of the function or its derivatives at discreet points $\{r_i\}$ in space. We show [Ref. 19] this in **Figure S1**, where for brevity, we have taken the function $f(x)$ to depend only one coordinate $x$.

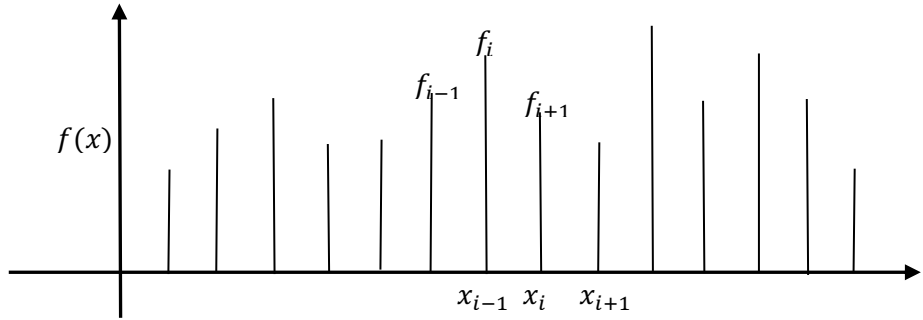

**Figure S1**. The values of a function f(x) at discreet point $\{x_i\}$ on the x-axis. A functional F[f] can be thought of as a multivariable function with the variable being the values $\{f_i\}$ of the function.

In applying the variational principle, we take the functional derivative of a functional and equate it to zero. By its definition, the functional derivative relates the first order change in the functional when its argument is varied by a small amount $\delta f(r)$. The functional derivative is denoted as $\frac{\delta F[f]}{\delta f(r)}$ and gives the first order change $\delta F$ in $F$ as

$$\delta F = \int \frac{\delta F[f]}{\delta f(r)}\delta f(r)\,dr \quad . \quad (S6)$$

Building up on the example of **Figure S1** and taking the functional $F(f_1,f_2\cdots;f_1',f_2'\cdots)$ to depend only the function and its derivative, the first order change in the functional can be written as

$$\delta F = \sum_i \left(\frac{\partial F}{\partial f_i}\delta f_i + \frac{\partial F}{\partial f_i'}\delta f_i'\right) \quad (S7)$$

This can be easily shown to lead to give

$$\frac{\delta F[f]}{\delta f(x)} = \frac{\partial g}{\partial f} - \frac{d}{dx}\left(\frac{\partial g}{\partial f'}\right) \qquad . \qquad (S8a)$$

This is easily generalized to three-dimensional situation. In that case the functional derivative is

$$\frac{\delta F[f]}{\delta f(\boldsymbol{r})} = \frac{\partial g}{\partial f} - \nabla \cdot \left(\frac{\partial g}{\partial \nabla f}\right) \qquad (S8b)$$

for functions $g(f, \nabla f, \nabla^2 f \cdots)$ in Eq. ($A2$) that depend on $f$ and $\nabla f$. In most of our work in this article, we will be dealing with such functional only. Take, for example, the kinetic energy functional $\frac{1}{2}\int |\nabla \varphi(\boldsymbol{r})|^2 d\boldsymbol{r}$ for a single particle in terms of its orbital $\varphi(\boldsymbol{r})$. Using the expression above, functional derivative of the kinetic energy functional with respect to $\varphi^*(\boldsymbol{r})$ will be $-\frac{1}{2}\nabla^2 \varphi(\boldsymbol{r})$.

There is a subtlety when derivatives are taken for a multivariable function; in light of the discussion above, the same applies to functional derivatives also. Depending on whether the variation $\delta f_i$ in Eq. (S7) is taken to be arbitrary or equal to $\epsilon h(x_i)$ for some chosen $h(x)$, where $\epsilon \to 0$, two different kinds of derivatives are defined for a function of multivariable. The former is known as the Fréchet derivative. In the latter case, the derivative is like the directional derivative in the direction of $h(x)$ and is called the Gâteaux derivative. It is evident that if the Fréchet derivative of a functional exists, its Gâteaux derivative will also exist. However, the converse is not necessarily true. The functionals we deal with are mostly Fréchet differentiable. We will therefore not dwell on this point any further and calculate the functional derivative using the formula

$$\frac{\delta F[f]}{\delta f(\boldsymbol{r})} = \lim_{\epsilon \to 0} \frac{F[f(\boldsymbol{r'}) + \epsilon \delta(\boldsymbol{r} - \boldsymbol{r'})] - F[f]}{\epsilon} \qquad . \qquad (S9)$$

When a function is to be found that minimizes or maximizes the value of a functional (or makes the functional stationary), the first order variation in the functional should vanish with respect to arbitrary small change in the function around the solution point. That means

$$\delta F = \int \frac{\delta F[f]}{\delta f(\boldsymbol{r})} \delta f(\boldsymbol{r}) d\boldsymbol{r} = 0 \qquad . \qquad (S10)$$

Since $\delta f(\boldsymbol{r})$ is arbitrary, condition ($A9$) in turn implies that the functional derivative

$$\frac{\delta F[f]}{\delta f(\boldsymbol{r})} = 0 \qquad (A11)$$

at the correct solution for $f(\boldsymbol{r})$. Eq. (S11) is thus the equation for finding function $f(\boldsymbol{r})$ that optimizes the value of the functional. However, often we also have the optimizing function satisfying certain condition. The condition could be expressed as a functional $C[f]$ being equal to zero. For example, if the function $f$ satisfies the normalization condition, it will be written as

$$C[f] = \int |f(\boldsymbol{r})|^2 d\boldsymbol{r} - 1 = 0 \qquad . \qquad (A12)$$

If $f$ satisfies such a condition, the correct solution for $f$ is found by using the Lagrange multiplier method. In this method, the variation of the functional $F[f] - \mu C[f]$ is made to vanish when $f(\boldsymbol{r})$ is changed by a small amount around the solution for it; here $\mu$ is a constant known as the Lagrange multiplier. Thus, the equation for $f(\boldsymbol{r})$ is now

$$\frac{\delta F[f]}{\delta f(\boldsymbol{r})} - \mu \frac{\delta C[f]}{\delta f(\boldsymbol{r})} = 0 \qquad . \qquad (A13)$$

The solution of this equation gives $f(\mathbf{r})$ that also depends on the Lagrange multiplier $\mu$. The value of $\mu$ is then fixed by the condition satisfied by $f(\mathbf{r})$. If there are more than one condition to be satisfied by $f(\mathbf{r})$, the new functional that is optimized is obtained by adding functional corresponding to each condition multiplied by a different Lagrange multiplier. Thus, there will be as many Lagrange multipliers as the number of conditions to be satisfied.

As an example, let us minimize the energy functional of Eq. ($S4$) with respect to $\varphi^*(\mathbf{r})$ under the condition that the wavefunction be normalized. In that case Eq. (S13) gives

$$-\frac{1}{2}\nabla^2\varphi + v_{ext}(\mathbf{r})\varphi - \mu\varphi = 0 \quad , \tag{A14}$$

which is the Schrödinger equation.

To summarize, in this supplemental material, we have given basic techniques of variational calculus that will suffice in carrying out the calculations performed in this article. Readers interested in more details are urged to consult Ref. 21.